\documentstyle[prd,aps,preprint,epsfig]{revtex} 
\tightenlines
\newcommand{\los}{\raisebox{-0.6ex}{$\scriptstyle 
\buildrel{\raisebox{-1pt}{$<$}}\over\sim$}} 
\newcommand{\gos}{\raisebox{-0.6ex}{$\scriptstyle 
\buildrel{\raisebox{-1pt}{$>$}}\over\sim$}} 
 
\title{Scaling solutions from interacting fluids} 
\author{Ana Nunes$^{1,2}$\thanks{email: anunes@lmc.fc.ul.pt},  \ Jos\'e P.Mimoso$^{1,3}$\thanks{email: jpmimoso@cii.fc.ul.pt} 
\& Tiago C. Charters$^{4,3}$\thanks{email: charters@cii.fc.ul.pt}} 
\address{$^{1}$Dep. F\'{\i}sica, Faculdade de Ci\^encias, Ed C8, Campo Grande\\ 
Universidade de Lisboa,  1749-016 
Lisboa, Portugal} 
\address{$^2$CMAF and $^3$CFNUL, Av. Prof. Gama Pinto, 2, 1649-003 Lisboa Codex, Portugal} 
\address{$^4$ Dep. Mec\^anica, Sec\c c\~ao Matem\'atica, ISEL\\ 
R. Conselheiro Em\' \i dio Navarro, 1, 1949-014 Lisboa, Portugal} 
\date{\today}

\begin{document}

\maketitle 
\begin{abstract} 
We examine the dynamical implications of 
an interaction between some of the fluid components of the 
universe. We consider the combination of three matter components, one of which is a perfect 
fluid and the other two are interacting. The interaction term generalizes the cases found in 
scalar field cosmologies with an exponential potential. We find that attracting scaling 
solutions are obtained in several regions of parameter space, that oscillating behaviour 
is possible, and that new curvature scaling solutions exist.  We also discuss the inflationary behaviour of the solutions and present some of the constraints on the strength of the coupling, namely those arising from nucleosynthesis. 
\end{abstract} 
 
\section{Introduction} 
There has been a growing appreciation of the importance of the asymptotic behaviour 
of cosmological models~\cite{Wainright + Ellis 97}. Indeed, unless there is a case for a cosmic 
coincidence~\cite{Zlatev+Wang+Steinhardt 99}, the features of the dynamics should be 
associated with some  stationary regime which should be obtained without fine-tuning of the 
initial conditions.  
 
A particular case which has attracted a great deal of interest concerns the possibility of 
obtaining cosmological scaling solutions with self-interacting scalar fields. It was shown  
that the exponential potentials yield the remarkable feature that the dynamics of the scalar 
field self-adjusts to that of matter so that the corresponding energy densities become 
proportional~\cite{Ratra+Peebles 88,Wetterich 88,WCL 93,Ferreira+Joyce 97,CLW 98}. 
In these solutions one envisages the interplay between the scalar field and matter, instead 
of focusing on the dynamics of models exclusively dominated by a self-interacting scalar  field 
as was the case in most of the models of inflation. 
It was shown that these solutions attract all the other phase space trajectories in the case 
of flat space models~\cite{CLW 98} and hence provide the late-time asymptotic behaviour for 
the scalar field cosmologies under consideration. This gives, on the one hand, a possible 
answer for why a non-vanishing scalar field does not introduce radical changes with respect to  
the usual Einstein-de Sitter rate of expansion of the universe. On other hand, it may 
contribute an explanation to the difference between the actual  density of matter and the 
critical energy density of the spatially flat isotropic models. Furthermore, the scalar field 
component would also fulfill the convenient role of delaying the time of matter-radiation 
equality which would help fitting the power spectrum of large scale 
structure~\cite{Coble et al 97,Wetterich 95,Viana+Liddle 98}.

So far the 
emphasis has been placed on the role of given families of potentials.  
In the literature we find mainly two sorts of potentials underlying the scaling behaviour: the 
exponential potentials and a class of power law 
potentials~\cite{Zlatev+Wang+Steinhardt 99,Ratra+Peebles 88,Liddle+Scherrer 99} respectively. 
However it is worth noticing that the solutions corresponding to the latter set of potentials 
were only shown to hold in the regime where the perfect fluid component fully dominates the 
expansion and the energy density of the scalar field is negligible. Regarding the solutions 
with an exponential potential the self-adjustment of the behaviour of the energy density of 
one or more scalar fields with that of matter has been investigated in 
Friedmann-Robertson-Walker (FRW) 
models~\cite{Ratra+Peebles 88,Wetterich 88,WCL 93,Ferreira+Joyce 97,CLW 98,Billyard+Coley+Hoogen 98,Hoogen+Coley+Wands 99} both with and without curvature, in spatially homogeneous, but anisotropic models~\cite{Coley+Ibanez+Hoogen 97}, and in FRW models in scalar-tensor gravity theories (also referred to as non-minimal coupling)~\cite{Billyard+Coley+Ibanez 98,Uzan 99,Amendola 99,Holden+Wands 00}. 
It was also shown by two of us~\cite{Nunes+Mimoso 00b} that every positive and monotonous 
potential which is asymptotically exponential yields a scaling solution as a global attractor. 
 
Recently Billyard and Coley~\cite{Billyard+Coley 00} have  
included an interaction term which transfers energy from the scalar field to the matter fields.  
In this work we analyse a related question, namely the role of the interaction between two of 
the components of the universe in promoting the scaling behaviour. We show that in the case of 
the self-interacting scalar field the self-tuning is a direct result of the energy transfer 
between the purely kinetic part and the vacuum-like part of the field accomplished by the 
gradient of the potential. We generalize this case by allowing for an interaction between two 
otherwise barotropic fluids and we examine its implications for the cosmological behaviour of 
homogeneous and isotropic universes. We consider a phenomenological interaction term between 
two of the matter components with the form 
$\propto H^\lambda \rho_1^\alpha \rho_2^\beta$, where $\alpha$, $\beta$ and 
$\lambda$ are constants which on dimensional grounds have to satisfy $\lambda+2(\alpha+\beta-1)=0$, 
and which, as we will show below, extends the case found in the scalar field with an 
exponential potential. Such a kind of energy exchange between two components  guarantees the 
global conservation of energy-momentum, that is, satisfies the contracted Bianchi identities 
and is akin to the type of dissipative terms considered in single component models where either 
a bulk viscosity term or matter creation are 
considered~\cite{Barrow 86,Lima+Germano 92,Zimdahl+Triginer+Pavon 96,Chimento+Jakubi+Pavon 99,Barrow 88}. 
We show that the consideration of such an interaction term yields a variety of situations where 
scaling solutions emerge. We produce a full classification of those cases where non-trivial scaling 
solutions (NTSS) are attractors. They include as particular case the scalar field cosmologies 
already referred to. Moreover we show that oscillating behaviour is also possible. In this latter 
case although the mean expansion rate is of a power-law type the various matter components 
oscillate.  
 
This paper is organized as follows. In the next section we review the scaling behaviour 
associated with scalar field cosmologies. We show how we may view this effect as the result of 
an interaction between the two limit situations where the energy of the scalar field lies in 
its kinetic part and the alternative case where it lies in its vacuum-like part associated with its 
potential. We show how the transfer of energy between these is promoted by the gradient of the 
potential. We also consider scalar field cosmological models within the extended framework of 
non-minimally coupled gravity theories. The Brans-Dicke (BD) scalar-tensor theory is remarkable 
in that  in the conformally transformed Einstein frame the original coupling between the BD 
scalar field and the curvature scalar of the space-time is traded into a re-defined scalar with an exponential 
potential which is now coupled to matter. In scalar-tensor theories there is then an intrinsic interaction between the 
conformally transformed scalar field energy density and the matter fields energy densities 
which lies at the root of the scaling behaviour of some solutions. In fact we consider the matter fields to be a 
combination of a perfect fluid and a radiation fluid and find a new type of scaling behaviour both in the Einstein and Jordan frames. 

In section 3 we generalize the 
scalar field models to a combination of three matter components where one of them is a 
perfect fluid and the other two are interacting. We perform a qualitative analysis which 
reveals that non-trivial scaling solutions arise in a number of situations and we classify them. 

 Finally in section 4 we summarize and discuss in the context of our general model a number of issues such as inflationary behaviour, curvature scaling solutions, bounds from nucleosynthesis and a phenomenological approach to the decay of massive particles out of equilibrium.

\section{Scalar field cosmologies} 
 
\subsection{Minimal-coupling theories} 
We consider a homogeneous and isotropic flat Friedmann-Robertson-Walker (FRW) universes filled 
with a perfect fluid, characterized by $p=(\gamma-1)\rho$, where $0\le \gamma \le 2$ is a 
dimensionless constant, and a self-interacting scalar field within the framework of Einstein's theory of general 
relativity. The field equations then read 
\begin{eqnarray} 
H^2 &=&  
\frac{\dot{\varphi}^2}{2} + V(\varphi) + \rho 
\label{e:Fried}\; , \\ 
\ddot{\varphi} &=& -3\frac{\dot{a}}{a}\,\dot{\varphi} 
-\frac{\partial V(\varphi)}{\partial \varphi} \label{e:sf1}\; , \\ 
\dot{\rho} &=& -3H \gamma \rho \label{e:pfcons} 
\; , 
\end{eqnarray} 
where the overdots stand for the derivatives with respect to the 
time $t$, and $H=\dot{a}/a$ (we use units in which $c=1=8\pi/m_p^2$, where 
$m_p=(\sqrt{G})^{-1}$ is the Planck mass and $G$ the gravitational constant). Another equation 
which is useful, albeit not independent of the former, is 
\begin{equation} 
\dot{H} = - \frac{1}{2}\,\left(\dot{\varphi}^2+\gamma \rho\right)\; . 
\end{equation} 
 
Defining $\rho_K={\dot{\varphi}^2}/{2}$ and $\rho_V = V(\varphi)$, the scalar field 
equation~(\ref{e:sf1}) may be cast as a system of two equations 
\begin{eqnarray} 
\dot{\rho_K} &=& -6H \rho_K - \sqrt{2} \left(\frac{V'}{V}\right) \rho_K^{1/2} \rho_V \label{sf_K}\\ 
\dot{\rho_V} &=& +\sqrt{2} \left(\frac{V'}{V}\right) \rho_K^{1/2} \rho_V \label{sf_V} \; , 
\end{eqnarray} 
which shows how the gradient of the potential promotes an  interaction between two limit 
``perfect fluids'', namely one associated with the scalar field's kinetic energy, $\rho_K$, 
that may be characterized by $\gamma_K=2$ and hence is akin to a stiff fluid, and the other 
associated with the potential energy, $\rho_V$, which has a vacuum like character and 
$\gamma_V=0$. In the absence of this interaction,  when $V$ is constant  (the case of an 
effective cosmological constant), there  is a smooth evolution from the early time domination 
of the perfect fluid with a higher value of $\gamma$  to the late time domination of the 
perfect fluid with the lower $\gamma$~\cite{Madsen et al 92,Coley+Wainwright 92}. In this case, there are no scaling 
solutions as the asymptotic solutions occur with the vanishing of one of the perfect fluid 
components. 
 
Let us  further define $\rho_c$ and $\rho_d$ as 
\begin{eqnarray} 
\rho_c=\rho_K + \rho_V &=& \frac{\dot{\varphi}^2}{2}+V(\varphi) \\ 
\rho_d=\rho_K - \rho_V&=& \frac{\dot{\varphi}^2}{2}-V(\varphi) \; . 
\end{eqnarray} 
which are, respectively, the energy density and the pressure of the  self-interacting scalar 
field~\cite{Madsen et al 92} in the co-moving observer frame. 
It then follows from Eqs.~(\ref{sf_K},\ref{sf_V})
\begin{eqnarray} 
\dot{\rho}_c &=& - 3H\,(\rho_c+\rho_d) \label{dotrc1}\\ 
\dot{\rho}_d &=& - 3H\,(\rho_c+\rho_d) - 2 V'(\varphi) \,\sqrt{\rho_c + \rho_d} \; . \label{dotrd1} 
\end{eqnarray} 
The second equation (\ref{dotrd1}) is the evolution equation for the 
scalar field pressure and we see that it is this one which explicitly involves the interaction 
between $\rho_1$ and $\rho_2$ as discussed above.

When we take $V(\varphi)$ to be an exponential potential, i.e., $V(\varphi)=V_0\, 
\exp{\left(\nu\varphi\right)}$, where $\nu$ is a constant, and introduce $\tau = \ln{a^3}$ as the new time variable, equations (\ref{dotrc1}) and (\ref{dotrd1}) become 
\begin{eqnarray} 
{\rho}_c' &=& - \,(\rho_c+\rho_d) \label{primerc}\\ 
{\rho}_d' &=& - (\rho_c+\rho_d) - \frac{\nu}{3H}\,(\rho_c-\rho_d)\,\sqrt{\rho_c+\rho_d} \; . \label{primerd} 
\end{eqnarray} 

We further introduce the new variables $x$ and $y$ defined as  
\begin{eqnarray} 
x&=&\frac{\rho_c}{3H^2}\label{def_x}\\ 
y&=&\frac{\rho_d}{3H^2}\label{def_y} 
\end{eqnarray} 
which correspond to the density parameters associated with $\rho_c$ and with $\rho_d$, 
respectively. Notice that in previous works on scaling solutions with exponential 
potentials~\cite{CLW 98,Billyard+Coley+Hoogen 98,Hoogen+Coley+Wands 99} it became popular 
to use expansion normalized variables which are square roots of density parameters. The only 
reason for that choice is to avoid the $\sqrt{\rho_K}$ which appears in the 
Eqs.~(\ref{sf_K},\ref{sf_V}). Since, in the next sections, we shall be interested in models 
where the interaction term will involve other powers  of the individual energy densities, 
apart from the power $1/2$, it is preferable to use the density parameters themselves given 
their immediate connection to observations. 
 
With the definitions (\ref{def_x}), (\ref{def_y})  we obtain 
\begin{eqnarray} 
x'&=& (\gamma-1)x-y+xy-(\gamma-1)x^2 \label{min_eqx}\\ 
y'&=& -  x +(\gamma-1)y +y^2-(\gamma-1)xy +\delta\, 
\sqrt{x+y}\,(x-y)  
\; , \label{min_eqy} 
\end{eqnarray} 
where $\delta=-\nu/\sqrt{3}$ and we have made use of the fact that the Friedmann constraint equation for the flat models, Eq.~(\ref{e:Fried}), 
yields $\Omega = 1-x-y$. 
 
It is straightforward to see that we just need to consider the  
triangle in the $x,y$ plane bounded 
by the invariant lines $x=y$, $x=-y$ and $x=1$. The vertices of this triangle are trivial 
critical points. The origin $(0,0)$ represents the vanishing of the scalar field energy density 
and hence of its pressure as well ($\rho_c=\rho_d=\rho_K=\rho_V=0$, the universe scales as 
$a\propto t^{2/(3\gamma)}$ and $\rho\propto a^{-3\gamma}$). The point $(1,1)$ represents the 
case where the kinetic energy of the scalar field dominates and thus corresponds to the 
massless field case ($\rho=\rho_V=0$, the universe scales as $a\propto t^{1/3}$ and 
$\rho_K\propto a^{-6}$). Finally,  the point $(1,-1)$ represents a spurious solution introduced 
by the change of variables we performed, for which $\rho=\rho_K =0$ and $\rho _V=3H^2$. Clearly, 
this is not a solution of the original 
equations (\ref{e:Fried},\ref{e:sf1},\ref{e:pfcons}). 
Notice though that in cases where the potential exhibits an underlying non-vanishing vacuum energy this critical point  
corresponds to the late-time domination of the vacuum energy and we have the well-known de Sitter exponential behaviour~\cite{Madsen et al 92,Wald 83}. 
 
Defining $\epsilon = \gamma-1$ it is easy to verify that  
there are critical points that correspond to $\bar{x}=1$ or $\bar{y}=\epsilon x$. The  first one, located at  ($\bar{x}=1,\bar{y}=\delta^2-1$), is a scalar field dominated solution ($\rho=0$) and only exists for 
$\delta^2/2<1$ which translates into $\nu^2<6$. This solution is an attractor and it 
corresponds to the well-known power-law inflationary solutions when 
$\nu^2<2$~\cite{Lucchin+Matarrese 85,Halliwell 87,Burd+Barrow 88,Coley+Ibanez+Hoogen 97}. The second one is given by  ($\bar{x}=(1+\epsilon)/\delta^2, 
\bar{y}= \epsilon(1+\epsilon)/\delta^2$) is the non-trivial scaling solutions found 
by \cite{Wetterich 88,Ratra+Peebles 88} and also studied by \cite{Ferreira+Joyce 97} and by \cite{CLW 98}. It 
only exists for  $\nu^2>3\gamma$ since $x<1$ \cite{Wetterich 88,CLW 98}. Linear stability analysis permits to distinguish two topological behaviours associated with this fixed point. It is a stable node if $\nu^2<24\gamma^2/(9\gamma-2)$ and a stable focus otherwise.

\subsection{Non-minimally coupled scalar fields} 
 
Scaling solutions in non-minimally coupled theories have also been found in the literature \cite{Billyard+Coley+Ibanez 98,Uzan 99,Amendola 99,Holden+Wands 00}. These theories, which can be formulated as general scalar-tensor gravity theories, are based on the Lagrangian~\cite{Bergmann:1968,Wagoner:1970,Nordtvedt:1970,Will 93}   
\begin{equation} 
L_\Phi =\Phi R -  
\frac{\omega(\Phi)}{\Phi}\, g^{ab} 
\Phi_{,a} \Phi^{,b} + 2  U(\Phi)+ 16 \pi L_m \quad , 
\label{eSTTactionc} 
\end{equation} 
where $R$ is the usual Ricci curvature scalar of a spacetime endowed with the 
metric $g_{ab}$,  $\Phi$ is a scalar field, $\omega(\Phi)$ is a dimensionless 
coupling function, $U(\Phi)$ is a function of $\Phi$, and ${\cal L}_m$ is the 
Lagrangian for the matter fields. They provide a most natural generalization of  
Einstein's general relativity (GR), and their investigation  enables a generic, 
model-independent approach to the main features and cosmological 
implications of the unification schemes which involve extra-dimensions.  
  
A distinctive feature of these theories is the coupling of the dynamical scalar field $\Phi $ 
to the scalar curvature $R$, which implies that the gravitational constant is now a function 
of $\Phi $, in fact $G=\Phi ^{-1}$. The archetypal of these theories is Brans-Dicke theory (BD), 
for which  $\omega (\Phi )$ is constant \cite{B+D 61}. 
The Lagrangian (\ref{eSTTactionc}) corresponds to  the so-called Jordan frame, in which the 
matter fields satisfy the equivalence principle (hence their energy-momentum tensor satisfies $\nabla_b T^{ab}=0$). This means that the $L_m$ terms do not 
explicitly involve the scalar field $\Phi $. By means of an appropriate conformal 
transformation of the space-time metric $g_{ab}$ to the so-called Einstein frame,  
\begin{equation} 
g_{ab} \to \tilde{g}_{ab} = (\Phi/\Phi_\ast) g_{ab}\; , 
\end{equation} 
where $\Phi_\ast$ is a constant allowing  the normalization of Newton's constant in the latter frame,  we recover a minimally coupled theory where the coupling of $\Phi $ to the curvature is traded into a coupling of the redefined 
scalar field with the matter fields. In fact $\Phi \to \varphi$~\cite{Dicke 62,Mimoso+Wands 95} such that  
\begin{equation} 
\frac{{\rm d}\ln{\Phi}}{{\rm d}\varphi} = \sqrt{\frac{16 \pi}{\Phi_\ast}} \, 
\alpha(\varphi)\, \label{def_alpha} 
 \; , 
\end{equation} 
where $\alpha=(\sqrt{2\omega(\varphi)+3})^{-1}$. The field equations for the flat FRW universes with a perfect fluid reduce then to the simple form~\cite{Mimoso+Nunes 98} 
\begin{eqnarray} 
3\frac{\dot{\tilde{a}}^2}{\tilde{a}^2} &=&  
\frac{8\pi}{\Phi_\ast}\,  \left[ \frac{\dot{\varphi}^2}{2} + \tilde{a}^{-3\gamma} \tilde{M}(\varphi) 
+ \, \tilde{V}(\varphi) \right]   
\label{eLPamV2} \\ 
\ddot{\varphi}&+&\frac{3\dot{\tilde{a}}}{\tilde{a}}\,\dot{\varphi}=-  
\tilde{a}^{-3\gamma} \frac{{\rm d}\tilde{M}(\varphi)}{{\rm d}\varphi}+  \frac{{\rm d}\tilde{V}(\varphi)}{{\rm d}\varphi}  \; , 
\label{eLPbmV2}\\ 
\dot{\tilde{\rho}}&=& - 3\frac{3\dot{\tilde{a}}}{\tilde{a}} \,\tilde{\rho} + \tilde{a}^{-3\gamma} \frac{{\rm d}\tilde{M}(\varphi)}{{\rm d}\varphi}\,\dot{\varphi}  
\label{eLPbmV3} 
\end{eqnarray} 
where the overdots stand for the derivatives with respect to the conformally transformed time 
$\tilde{t}$, 
$\tilde{V}(\varphi)=\Phi_\ast^2\, U(\Phi(\varphi))/(8\pi\Phi^2(\varphi))$, 
$\tilde{M}(\varphi)=\mu \,(\Phi(\varphi)/\Phi_\ast)^{-(2-3\gamma/2)}$ 
and  $\mu\,$ is the constant defined by 
 $\mu\equiv \rho a^{3\gamma}\,$ which fixes the initial 
conditions for the energy-density of the perfect fluid (note that for the sake of making clear the effect of the conformal transformation on the coupling  we have written the latter equations with the $8\pi/\Phi_\ast$ in spite of of our choice of units; in what follows we shall again set it equal to 1).
From Eq.~(\ref{eLPbmV2}) and the definition of $M(\varphi)$ it is apparent that if the matter sources are radiation fields (the case in which $\tilde{M}=\mu$ is constant) or, alternatively,  vacuum (the case in which $\mu=0$ and hence $\tilde{M}=0$) the 
coupling vanishes, which translates the fact that the latter cases are conformally invariant. 
 
It is a simple matter to see from Eq.~(\ref{def_alpha}) that 
in the BD case, where $\omega$ (and hence $\alpha$) is constant,
$\Phi\propto \exp{\left(\sqrt{2}\,\alpha \varphi \right)}$ so that $\tilde{M}$ is  exponential and so is $\tilde{V}$ if the original potential $U(\Phi)$ is a power-law in $\Phi$. This means that the coupling encapsulated in the $M(\varphi)$ function amounts in this case to a modification of the constant coefficients of the dynamical system associated with the general relativistic case yielding scaling solutions for suitable values of the parameters~\cite{Billyard+Coley+Ibanez 98,Uzan 99,Amendola 99,Holden+Wands 00}.  
  
Here we shall restrict ourselves to the original BD theory where, besides having a constant  $\omega$ (or equivalently $\alpha$), we also have a vanishing $U(\Phi)$ in Eq.~(\ref{eSTTactionc}). We will continue to focus on the flat Friedmann-Robertson-Walker model, but instead of just having a  perfect fluid we will also consider the simultaneous presence of radiation (hereafter we shall drop the tildes referring to the conformal frame quantities). Denoting the 
perfect fluid (respectively, radiation) energy density and pressure by $\rho_m$ and 
$p_m$ (respectively, $\rho_r$ and $p_r$), and the energy density and the 
pressure of the redefined scalar field $\varphi $ by $\rho_\varphi$ and $p_\varphi$, the field 
equations are 
\begin{eqnarray} 
3H^2 &=& \, {\rho_r+\rho_\varphi+\rho_m} \, \; ,\\ 
\frac{{\rm d}{H}}{{\rm d}{t}}  &=& 
 - \frac{1}{2} \, \left((\rho_r+p_r)+(\rho_\varphi+p_\varphi)+(\rho_m+p_m) 
\right) 
             \; , 
\end{eqnarray} 
and the evolution equations for the energy densities of the three fluids are 
given by 
\begin{eqnarray} 
\frac{{\rm d}{\rho_r}}{{\rm d}{t}} & = & -4H\,\rho_r\\ 
\frac{{\rm d}\rho_m}{{\rm d}{t}} & = & -3H(\rho_m+p_m) 
    +\frac{1}{ \sqrt{2\omega+3}}\, 
        (3p_m-\rho_m) \frac{1}{2}\frac{{\rm d}\varphi}{{\rm d}{t}} \; , \\ 
\frac{{\rm d}\rho_\varphi}{{\rm d}{t}} & = & -6H\rho_\varphi 
    -\frac{1}{ \sqrt{2\omega+3}}\,  
        (3p_m-\rho_m) \frac{1}{2}\frac{{\rm d}\varphi}{{\rm d}{t}} 
\; , 
\end{eqnarray} 
rendering explicit the interaction that now leads to a transfer of energy between the matter  
perfect fluid and the stiff fluid ($\dot{\varphi}^2/2$) associated with the massless scalar field $\varphi$. 
Since $p_m=(\gamma-1)\rho_m$ and given the definitions of $\rho_m$ and of 
$\rho_\varphi$, we see that the interaction term 
has the same form as in the case of the minimally coupled scalar field with an exponential potential. 
Indeed, we may write for each component an equation of state of the form 
$p_m=(\gamma-1)\rho_m - p_{int}$ and $p_\varphi=\rho_\varphi + p_{int}$, where the 
interaction term can be cast  as 
\begin{equation} 
p_{int}=  (3H)^{-1} \frac{1}{ \sqrt{2\omega+3}}\,  
        (4-3\gamma)\rho_m \sqrt{\rho_\varphi} \; .  
\end{equation} 
There are two major differences with regard to the minimally coupled case. On one hand, the dimensionless constant $\delta$ that measures the strength of the interaction depends on $\gamma$ according with 
\begin{equation} 
\delta=\frac{1}{\sqrt{3}}\,\frac{4-3\gamma}{2-\gamma}\,\alpha \; .
\end{equation} 
On the other hand, the $\gamma$-index of the non-interacting fluid is now fixed to take the value $4/3$ of radiation and $\gamma_{1}$  remains equal to 2, and $\gamma_{2}$ now takes the value of the free parameter $\gamma$. 
Therefore taking into account these differences the dynamical system~(\ref{min_eqx},\ref{min_eqy}) adopts the form  
\begin{eqnarray} 
x'&=& \frac{(2-3\gamma)}{3(2-\gamma)}\,x-y+xy-\frac{(2-3\gamma)}{3(2-\gamma)}\,x^2 \label{nmin_eqx}\\ 
y'&=& -  x +\frac{(2-3\gamma)}{3(2-\gamma)}\,y +y^2-\frac{(2-3\gamma)}{3(2-\gamma)}\,xy +\delta\, 
\sqrt{x+y}\,(x-y)  
\; . \label{nmin_eqy} 
\end{eqnarray} 
It is a simple matter to see that as before we obtain two NTSS. One at the point $(\bar{x}=1,\bar{y}=\delta^2/2-1)$ which corresponds to a late-time domination of the coupled components, the scalar field and the  
$\gamma$-fluid. Radiation is depleted and the universe scales as power-law $a\propto t^{2/(3\Gamma)}$ where $\Gamma=\gamma+\delta^2/2\,(1-\gamma/2)$ is defined by the position of the fixed point. Another NTSS exists at the point $(\bar{x}= 2/(\sqrt{3}\delta \alpha),\bar{y} = (2-3\gamma)/(3(2-\gamma)) \,\bar{x}$ where the three components are simultaneously present and the energy densities of the two that are linked by the interaction have adjusted to the radiation behaviour. Thus the scale factor evolves as  
$a\propto t^{1/2}$ as in a universe dominated by radiation, but with the major difference that both the BD-scalar fluid and the  
$\gamma$-fluid remain in relevant proportions (this scaling solution corresponds in fact to the $C_{RM}$ fixed point in the paper of Amendola~\cite{Amendola 00}). 
 
If we reverse the conformal transformation we are able to recover the exact solutions in the Jordan frame (which is commonly taken to be the physical frame). The case of the fixed point on the line $x=1$ corresponds to the power-law solutions found by Nariai~\cite{Nariai 68} and includes the dust solution ($\gamma=1$) derived by Brans and Dicke~\cite{B+D 61}. The other non-trivial scaling solution corresponds again to a power-law in the Jordan frame, but differs from the single-fluid exact solutions found in the literature~\cite{Mimoso 93,Barrow+Mimoso 94}. For this we have  
\begin{eqnarray} 
a(t) &\propto & t^{\pm \sqrt{2}/2} \\ 
\Phi(t) &\propto & t^{\pm 2(1-\sqrt{2})} \; ,  
\end{eqnarray}  
where $a$ and $t$ are the scale factor and the time coordinate in the Jordan frame. This exact power-law solution is new and it differs from the corresponding behaviour of the radiation BD models for which 
$a\propto t^{1/2}$.  
 
\section{Cosmologies with interacting fluids} 
 
In this section we shall construct a general model which includes the two previous classes of 
models as particular cases. We assume the matter content of the universe to be the 
combination of three components. One that evolves without interaction with the other two and which  satisfies the usual barotropic $p=(\gamma-1)\rho$ law (in the particular case of the non-minimal coupling models of the last sub-section it would correspond to the radiation component). The two remaining components are mutually 
coupled by an interaction term of the type
\begin{equation}
p_i = (\gamma_i-1) 
\,\rho_i \pm \eta \,H^\lambda \rho_i^\alpha\, \rho_j^\beta
\; , \label{def_int}
\end{equation}
where $i,j=1,2$, and $\alpha$, $\beta$, $\lambda$ are constants, and $\eta$ is a dimensionless constant. The $\pm$ sign of the interaction term means that if it takes one of the signs for one of the components it necessarily takes the opposite sign for the other, thus ensuring the overall conservation of the energy-momentum tensor $\nabla_b (T_{(1)}^{ab}+T_{(2)}^{ab})=0$ as required by the Bianchi contracted identities. Moreover, using dimensional considerations, it is a simple matter to verify that $\alpha$, $\beta$ and 
$\lambda$  have to satisfy $\lambda+2(\alpha+\beta-1)=0$ for the interaction to be a pressure. 
The dependence of the interaction term on the products of  powers of the densities of the particle species reflects, on one hand, the fact that one expects that it should be proportional to the number of collisions between the particles of each species and thus on the product of their number densities, and, on the other hand, the fact that in the collisions the energies of the particles  will be shifted. Furthermore the factor depending on $H$ accounts for the characteristic time of the individual interactions. Since a detailed relativistic kinetic model of the interaction between two fluids is missing, we keep our model as general as possible by using the two free parameters of the set $\alpha$, $\beta$ and $\lambda$ to allow for unknown aspects of the interaction. The model under consideration aims at a phenomenological description of transient phases of the universe which arise when some of the material components are not in thermal equilibrium. One must be wary that this is in fact a likely situation as the expansion of the universe is permanently trying to pull the matter fields out of equilibrium~\cite{Barrow 86,Bernstein 88}. In this sense the type of interaction generalizes those found in the literature when  viscous pressure and matter creation terms are considered in the literature on dissipative isotropic  
models~\cite{Lima+Germano 92,Zimdahl+Triginer+Pavon 96,Chimento+Jakubi+Pavon 99,Barrow 86,Barrow 88}.  
Since our main purpose is to produce a classification of the cases which lead to scaling solutions, where two or more components self-adjust their behaviour so that their energy-densities scale with  the same rate, we shall be essentially  concerned with the dynamical aspects of this model, rather than with the specifics of particular models. Nevertheless we emphasize that the post-inflationary reheating period, the decay of massive particle species into lighter ones and the general situations of decoupling of particle species provide examples of the situations where the present model may be applied. We will address some of these examples in the section IV. 
  
We shall denote the energy density of the first component $\rho_X$ and assume 
that it has a barotropic equation of state characterized by a constant adiabatic index 
$\gamma_X$ which we shall leave as a free parameter. 
 
The field equations read 
\begin{eqnarray} 
3H^2&=&\rho_X +\rho_1+\rho_2  \label{Fried_if1}\\ 
\dot{H}&=& -\frac{1}{2}\left[(\rho_X+p_X)+(\rho_1+p_1)+(\rho_2+p_2)\right] \label{dotH_if1}\\ 
\dot{\rho}_X &=&-3H \,(\rho_X+p_X) \label{dotrhoX_if1}\\ 
\sum_i \dot{\rho}_i  &=&-3H \,\sum_i (\rho_i+p_i)  \; . 
\end{eqnarray} 
Here $\rho_j$, $p_j$, with 
$j=1,2$ are, respectively, the energy density and the pressure of the $j^{\rm th}$ matter 
component measured by a comoving observer. 
 
We define 
\begin{eqnarray} 
\rho_c&=& \rho_1+\rho_2 \\ 
\rho_d&=& \rho_1-\rho_2\\ 
{{\gamma}}_c &=& \frac{\gamma_1+\gamma_2}{2}\\ 
{{\gamma}}_d &=& \frac{\gamma_1-\gamma_2}{2} 
\end{eqnarray} 
so that the field equations become 
\begin{eqnarray} 
3H^2&=&\rho_X +\rho_c \label{ef}\\ 
\dot{H}&=& -\frac{1}{2}\left[\gamma_X \rho_X+{{\gamma}}_c\,\rho_c+ 
{{\gamma}}_d\,\rho_d\right] \label{eH}\\ 
\dot{\rho}_X &=&-3H \,\gamma_X\,\rho_X\label{eX}\\ 
\dot{\rho}_c &=&-3H \,\left({{\gamma}}_c\,\rho_c+{{\gamma}}_d\, 
\rho_d  \right) \; \label{erc}\\ 
\dot{\rho}_d &=&-3H \,\left({{\gamma}}_d\,\rho_c+{{\gamma}}_c\, 
\rho_d  \right) + 3\eta H^{\lambda +1}2^{1-\alpha -\beta}(\rho _c+\rho _d)^{\alpha}(\rho _c-\rho _d)^{\beta}\;. \label{erd} 
\end{eqnarray} 

It is also convenient to introduce the density parameters 
\begin{eqnarray} 
x&=&\frac{\rho_c}{3H^2}\label{ex}\\ 
y&=&\frac{\rho_d}{3H^2}\label{ey} 
\end{eqnarray} 
as dynamical variables and the new time variable $\tau = \ln{(a/a_o)^3}$. Then, from Eq.~(\ref{ef}), $\rho_X/3H^2=1-x$ and 
the field equations reduce to the following dynamical system 
\begin{eqnarray} 
x'&=& -\left(({{\gamma}}_c -\gamma_X)x  +{{\gamma}}_d y\right)\, 
(x-1) \label{ex'}\\ 
y'&=& - \gamma _d x - \gamma _c y + y\left(\gamma_X(1-x) + \gamma _c x + \gamma _d y\right) +  \left(\frac{2}{3}\right)^{1-\alpha-\beta}\,\eta 
\,H^{\lambda +2\alpha +2\beta -2}\,(x+y)^\alpha\,(x-y)^\beta \;,\label{ey'} 
\end{eqnarray} 
where the prime denotes differentiation with respect to the 
$\tau$ variable. 
 
It is remarkable that the dimensional relation that must be satisfied by the parameters, $\lambda+2(\alpha+\beta-1)=0$, is precisely  the one that renders the above system  
autonomous. From (\ref{ef}),  (\ref{ex}) and  (\ref{ey}), we have the 
restrictions $0 \leq x \leq 1$, $|y| \leq x$ which define the 
phase space of the system, a triangle in the $x,y$ plane bounded 
by the invariant lines $x=y$, $x=-y$ and $x=1$. Using the 
symmetry $(y, \gamma _d, \alpha , \beta ) \rightarrow (-y, 
-\gamma _d, \beta ,\alpha )$ we may restrict ourselves to the 
when $\gamma _d >0$, i.e., $\gamma _1 > \gamma _2$. Rescaling 
time through the factor $\gamma _d$ and defining the new 
parameters $\displaystyle{ \epsilon =\frac {\gamma _X-\gamma _c}{\gamma _d}}$, 
$\displaystyle{\delta = \frac {\eta (3/2)^{\alpha +\beta -1}}{\gamma _d}}$, 
equations (\ref{ex'}) and  (\ref{ey'}) may be simplified and rewritten in a more compact form as 
\begin{eqnarray} 
\dot x & = & (1-x)(-y+\epsilon x) \nonumber \\ 
\dot y & = & -x +\epsilon y + y(y-\epsilon x) + \delta 
(x+y)^{\alpha }(x-y)^{\beta }. \label{3.1} 
\end{eqnarray} 
 
The parameter $\delta $ measures the strength of the interaction, and $|\epsilon | < 1$ 
indicates that $\gamma _X$ lies within the interval $(\gamma _1, \gamma _2)$.  The cases of 
the previous section correspond to $\alpha =1/2$, $\beta =1$. In the minimally coupled scalar 
field cosmologies, $\epsilon = \gamma -1$, whereas in the BD case 
$\epsilon = \frac {2-3\gamma }{3(\gamma -2)}$. 
 
The points $(0,0)$, $(1,1)$ and $(1,-1)$ are always equilibrium 
points for this system. They correspond to the trivial scaling 
solutions which exist when $\eta =0$. We shall say that an 
equilibrium point of (\ref{3.1}) is a {\em non trivial scaling 
solution} (NTSS) if it is none of the above and furthermore it is 
stable. Our study of system (\ref{3.1}) will be directed towards 
the search for non trivial scaling solutions. The additional 
symmetry $(y, t, \epsilon, \alpha , \beta ) \rightarrow (-y, -t, 
-\epsilon, \beta ,\alpha )$ is useful to reduce the number of 
different possible cases. The techniques we  employ are the standard methods of 
qualitative theory of planar systems, assisted, in the degenerate cases,  by numerical 
integration. 
 
\subsection{Dynamics on $x=1$} 
 
When $\rho _X=0$, the interplay between the two interacting fluids 
is described by the single equation 
 
\begin{equation} 
\dot y = (y^2 -1) + \delta (1+y)^{\alpha }(1-y)^{\beta }. 
\end{equation} 
 
It is straightforward to check that NTSS on $x=1$ show up in the 
three cases depicted in Figure 1. Denoting generically by ($\bar{x},\bar{y}$) the coordinates of any of these fixed points, we see from Eqs. (\ref{ef}), (\ref{erc}), (\ref{ex}) and  (\ref{ey}) that the scaling solution is characterized by  $a\propto t^{2/(3\Gamma)}$ where $\Gamma = \gamma_c + \gamma_d\, \bar{y}$.

\subsection{Global dynamics for $|\epsilon |\geq 1$} 
 
When $|\epsilon | \geq 1$ there are no equilibria of (\ref{3.1}) 
outside the boundary of the phase space. For $\epsilon < -1$, 
i.e., $\gamma _X < \gamma _2 < \gamma _1$, the origin is a global 
attractor. For $\epsilon >1$, the line $x=1$ becomes the global 
attractor, so that the stable equilibria described in the 
preceding subsection become NTSS for the full system, which behaves 
as shown in Figure 2. \vskip 1cm

When the non-interacting  matter component behaves as one of the two interacting fluids, we 
have the degenerate cases $\epsilon =1$ ($\gamma _X = \gamma _1$) 
and $\epsilon =-1$ ($\gamma _X = \gamma _2$). These cases also 
provide NTSS on the lines $y=x$ and $y=-x$, respectively, as shown 
in Figure 3 for the cases when other equilibria on $x=1$ coexist. 
In the cases of Figures 3.a) and 3.b), that is, when $\gamma _X = 
\gamma _2$  and $\alpha >1$,  $\rho _X$  would have been 
negligible in the past, while in the case of Figure 3.c), that is 
when $\gamma _X = \gamma _1$  and $\beta <1$, $\rho _X$ would 
have dominated in the past.

\subsection{The case  $|\epsilon | < 1$ and $\alpha, \beta >1$} 
 
When $(1+\epsilon )^{\alpha -1}(1-\epsilon )^{\beta -1} \leq 
1/\delta $, there are no equilibria in the interior of the phase 
space, and the dynamics is either trivial or as shown in Figure 
4.a) according to whether $ (\frac {2}{\alpha +\beta -2} 
 )^{\alpha +\beta -2}(\alpha -1)^{\alpha -1} 
(\beta -1)^{\beta -1}$ is smaller or greater than $1/\delta $, 
respectively. When $(1+\epsilon )^{\alpha -1}(1-\epsilon )^{\beta 
-1} > 1/\delta $ we have always $ (\frac {2}{\alpha +\beta -2} 
 )^{\alpha +\beta -2}(\alpha -1)^{\alpha -1} 
(\beta -1)^{\beta -1} > 1/\delta $, since the first member of the 
preceding inequality is precisely the maximum of  $(1+\epsilon 
)^{\alpha -1}(1-\epsilon )^{\beta -1}$ as a function of $\epsilon 
$, attained at $\epsilon _M = \frac {\alpha - \beta } {\alpha 
+\beta -2}$. In this case, there exists another equilibrium in the 
interior of the phase space, whose stability is determined by the 
sign of $\epsilon _M - \epsilon $. Thus, we have an additional NTSS 
 in the cases of Figure 4.b) and 4.c). 
The coordinates of the new solution are $(\bar x, \epsilon \bar 
x)$, where $\bar x $ satisfies 
\begin{equation} 
1/\delta = \bar x^{\alpha +\beta -1} (1+\epsilon )^{\alpha 
-1}(1-\epsilon )^{\beta -1} .\label{3.barx} 
\end{equation} 
At this scaling solution and, in fact, at every scaling solution in the interior of the phase space domain, 
the universe scales as $a\propto t^{2/(3\gamma_X)}$.

 In the case of Figure 4.b), we have 
the new feature of a neutrally stable NTSS. In the case of Figure 
4.b), the basin of attraction of the NTSS shares the phase space 
with the basin of attraction of $(1,-1)$.

\subsection{The case  $|\epsilon | < 1$, $\alpha, \beta <1$ and $\alpha + 
\beta \geq 1$} 
 
As above, an equilibrium in the interior of the phase space exists 
only when  $(1+\epsilon )^{1-\alpha }(1-\epsilon )^{1-\beta } < 
\delta $. For $(1+\epsilon )^{1-\alpha }(1-\epsilon )^{1-\beta } 
\geq \delta $ we have always $ (\frac {2}{\alpha +\beta -2} 
 )^{\alpha +\beta -2}(\alpha -1)^{\alpha -1} 
(\beta -1)^{\beta -1} > \delta $ since, as before, this is the 
maximum value of the function $(1+\epsilon )^{1-\alpha 
}(1-\epsilon )^{1-\beta }$, attained at $\epsilon _M = (\beta 
-\alpha )/(2-\alpha -\beta )$. The dynamics in this case is as 
shown in Figure 5.a). With respect to Figure 2.c), the change in 
$\epsilon $ has turned the NTSS into the only attractor. 
 
When $(1+\epsilon )^{1-\alpha }(1-\epsilon )^{1-\beta } < \delta 
$, we may have either $(1+\epsilon _M)^{1-\alpha }(1-\epsilon 
_M)^{1-\beta } \leq \delta $ or $(1+\epsilon _M)^{1-\alpha 
}(1-\epsilon _M)^{1-\beta } > \delta $. In the first case, there 
are no NTSS on the boundary of the phase space, and we have at 
$(\bar x, \epsilon \bar x)$, $\bar x$ given by (\ref{3.barx}) a 
neutrally stable (resp. asymptotically stable) NTSS when $\epsilon 
= \epsilon _M$ (resp. $\epsilon < \epsilon _M$), see Figures 5.b) 
and 5.c). Notice that, in both situations, $(0,0)$ is an attractor 
for a set of positive measure of initial conditions. In the second 
case, the dynamics is as shown in Figures 5.d) and 5.e), according 
to whether $\epsilon < \epsilon _M$ or $\epsilon > \epsilon _M$. 
In both situations, the NTSS is a global attractor.

\subsection{The case  $|\epsilon | < 1$, $\alpha, \beta <1$ and $\alpha + 
\beta < 1$} 
 
In this case, we shall have an equilibrium in the interior of the 
phase space at $(\bar x, \epsilon \bar x$ with $\bar x $ given by 
(\ref{3.barx}) whenever  $(1+\epsilon )^{1-\alpha }(1-\epsilon 
)^{1-\beta } > \delta$. When this happens, there are always 
equilibria on the line $x=1$, which falls into the case of Figure 
1.c), and the additional equilibrium is a saddle. The dynamics is 
shown in Figure 6.a). With respect to Figure 2.c), the NTSS 
survives as a partial attractor. 
 
When $(1+\epsilon )^{1-\alpha }(1-\epsilon )^{1-\beta } \leq 
\delta$ there are no equilibria in the interior of the phase space 
and we may still have either $ (\frac {2}{2-\alpha -\beta } 
 )^{2-\alpha -\beta }(1-\alpha )^{1-\alpha } 
(1-\beta )^{1-\beta } > \delta $ or $ (\frac {2}{2-\alpha -\beta } 
 )^{2-\alpha -\beta }(1-\alpha )^{1-\alpha } 
(1-\beta )^{1-\beta } \leq \delta $. Only the first case may 
provide NTSS, when, moreover, $\epsilon 
>0$, and the dynamics is then as shown in Figure 6.b). With respect to Figure 2.c), 
the dynamics is only slightly changed, and the NTSS.

\subsection{The case  $|\epsilon | < 1$ and  $\alpha \leq 1 < \beta $ or 
$\alpha < 1 \leq \beta $} 
 
An equilibrium in the interior of the phase space exists if and 
only if $(1+\epsilon )^{\alpha -1}(1-\epsilon )^{\beta -1} > 1/ 
\delta $. In this case, it is easy to check that the equilibrium 
on $x=1$ is a saddle, and that the additional equilibrium is a 
sink and hence a NTSS. In the complementary case, there can only be 
non trivial scaling solutions when the dynamics on  $x=1$ is as in 
Figure 1.b). Then, the NTSS on $x=1$ is a sink for the full system, 
and a global attractor. The dynamics is represented on Figure 7. 
Finally, it is easy to check using the symmetries that the case 
$\beta \leq 1 < \alpha $ provides no NTSS.

\subsection{The case  $|\epsilon | < 1$, $\alpha = \beta =1$ and $\delta  >1$} 
 
In this case, we have also a neutrally stable NTSS in the interior 
of the phase space. The dynamics is shown in Figure 8. With 
respect to Figure 5.b), the behaviour is similar, except that 
$(0,0)$ is now a saddle so that, in the present case, there are no 
attractors.

\section {Discussion and conclusions} 
In this work we have produced a qualitative study of the dynamics of the flat FRW model with three matter components one of which is barotropic and the other two are coupled, exchanging energy. The consideration  of the flat model is motivated by observational data that indicate that the universe is flat~\cite{White et al 93,Perlmut et al 98,Efstathiou et al 99}. 
Since these results also suggest that the density parameter of matter (including baryonic matter and cold dark matter) only accounts for $\Omega_m \simeq 0.35$, some additional form of dark energy should also intervene. Irrespective of the particular proportions taken by each of the major material components the mere fact that any other components might be present apart from the usual barotropic radiation and matter fluids of the standard cosmological model~\cite{Weinberg 72,Kolb+Turner 90} raises the worry that   
the usual expansion rates during the radiation and Einstein-de Sitter epochs of the standard model would be significantly disrupted. Attracting scaling solutions  provide a possible way of reconciling the presence of several matter components with the standard model as the various energy densities at stake all scale with the same rate as that of the barotropic fluid if the latter is non-vanishing. For the general type of interaction we considered in this paper, we have presented all the classes of parameter values 
for which system (\ref{3.1}) exhibits non trivial scaling 
solutions. Those which correspond to equilibria outside the line 
$x=1$, that is, for which $\rho _X\neq 0$, are especially 
interesting from the physical point of view. These are the cases 
of Figures 3, 4.b) and c), 5.b), c), d) and e), 7.a) and 8. Among 
these, the cases of Figures 4.b), 5.b) and 8 are also remarkable. 
They correspond to situations where for a set of positive (or even 
full) measure of initial conditions in phase space the solutions 
are periodic, and so the relative abundance of each of the three 
fluids oscillates. 
We would also like to remark that this class of models opens the possibility of having simultaneously scaling and periods of inflationary behaviour.  
 
\subsection{Inflation\label{ssec_inflation}} 
In fact, it is easy to see that the condition for inflationary behaviour ($\ddot{a}>0$) is 
\begin{equation} 
y < \epsilon x + \frac {2 - 3\gamma _X}{3\gamma _d}, 
\end{equation} 
and thus defines a half-plane in the $x,y$ plane. This condition does not depend on the values of the parameters $\lambda$, $\alpha$, $\beta$ and $\delta$, it just depends on $\gamma_X$, $\gamma_1$ and $\gamma_2$, and adjusting  the latter 
the triangle of the physical phase space may have partial or full overlap with the inflationary half-plane.   
When the boundary line $y_\ast(x) = \epsilon x + 
(2 - 3\gamma _X)/(3\gamma _d)$ lies in the $y>0$ half-plane and intersects the $x=1$ vertical line above the point  $(1,1)$  
all the trajectories are inflationary. This situation arises when $\gamma_X<2/3$ and $\gamma_1<2/3$ (recall that we are assuming $\gamma_2<\gamma_1$). The opposite situation arises when the boundary line $y_\ast(x)$ lies in the $y<0$ half-plane and intersects the $x=1$ vertical line below the point  
$(1,-1)$ in which case no trajectory is inflationary. This happens for $\gamma_X>2/3$ and $\gamma_2>2/3$. In the intermediate situations, namely when $\gamma_2<2/3<\gamma_1$, the inflationary region corresponds to the portion of the phase space triangle below the $y_\ast(x)$ line. The intersections of this line with the triangle are at the points  
\[(1,\frac{2-3\gamma_c}{3\gamma_d}) \; , \] 
\[(\frac{1}{1-\epsilon}\,\frac{2-3\gamma_X}{3\gamma_d},\frac{1}{1-\epsilon}\,\frac{2-3\gamma_X}{3\gamma_d}) \qquad\quad {\rm when} \quad \gamma_X<\frac{2}{3} \; ,\] 
\[(\frac{1}{1+\epsilon}\,\frac{3\gamma_X-2}{3\gamma_d},\frac{1}{1+\epsilon}\,\frac{2-3\gamma_X}{3\gamma_d}) \qquad\quad  {\rm when} \quad \gamma_X>\frac{2}{3}  \; .\] 
 
Let us consider what happens in the scalar field models previously discussed in Section II. 
In  the minimal coupling case 
the inflationary solutions lie below the $y_\ast(x) =(\gamma-1)x+(2-3\gamma)/3$ line. This line intersects the $x=1$ frontier of the triangle at $y=-1/3$ regardless of its slope defined by the value of $\gamma$. For $\gamma=1$, that is, for dust, the line is horizontal and the inflationary region is a  triangle. As we consider smaller values of $\gamma$, the slope of the line $y_\ast(x)$ becomes increasingly negative and in the limit case of $\gamma=0$ it becomes $-1$. Notice that this limit value of $\gamma$ does not yield NTSS solutions in the interior of the phase-space domain. It corresponds to a cosmological constant and it is equivalent to having a non-vanishing vacuum energy in the exponential potential. The converse happens as we consider values of $\gamma >1$. The slope increases up to a maximum value when  the limit $\gamma=2$ is chosen and the boundary line is parallel to $y=x$. So there is always an inflationary region in the phase diagram of the models and this means that portions of the trajectories approaching the attractors will exhibit inflationary transients. Let us consider now the question of   whether the scaling solutions fall within those regions. In the $\nu^2<3\gamma<6$ case, the only fixed point lies on the $x=1$ vertical line. When $\nu^2<2$ its $y$ coordinate satisfies $y<-1/3$ and hence falls within the inflationary region. 
In the $\nu^2>3\gamma$ case, the stable scaling solution $\bar{x}=3\gamma/\nu^2$, $\bar{y}=3\gamma(\gamma-1)/\nu^2$ belongs to the inflationary region only if $\gamma<2/3$.  
This means that we may have NTSS which exhibit inflationary behaviour. However, as the most interesting models from the viewpoint of the late time behaviour of the universe are those for which the perfect fluid has $\gamma \ge 1$, namely $\gamma =1, 4/3$, the remarkable issue is that a non-negligible set of solutions naturally undergo a finite period of inflation before reaching the attractor.  
 
The models leading to oscillatory behaviour are also  interesting in what concerns inflation. Consider the model associated with the figure 8, that is, the model characterized by $|\epsilon|<1$, $\alpha=\beta=1$ and $\delta>1$ . The NTSS is located at $\bar{x}=1/\delta$, $\bar{y}=\epsilon/\delta$ and it corresponds to an expansion that tracks the non-interacting perfect fluid so that $a\propto t^{2/(3\gamma_X)}$. Therefore it is immediate to see that this NTSS falls within the inflationary region of the phase-diagram if $\gamma_X <2/3$ and off it otherwise. Taking into consideration what was expounded at the beginning of this subsection, in order for the inflationary half-plane to overlap the phase-plane triangle one requires that $\gamma_2$ be smaller than $2/3$. Assuming $\gamma_X>2/3$ and $\gamma_2<2/3$ we have then that the oscillatory trajectories beyond certain radius from the fixed point will undergo cyclic periods of inflation since they periodically cross the inflationary region of the phase-plane. 
Notice that since the model under consideration corresponds to a non-linear oscillator the periods associated with the trajectories increase from the immediate (and extremely small) neighbourhood of the fixed point, where the linear approximation holds true and $T\simeq 2\pi/\epsilon(1-1/\delta)\sqrt{(1-\epsilon^2)/(\delta-1)}$, to the regions close to the limits of the triangle where the period becomes infinite.

\subsection{Curvature scaling solutions} 
An interesting case which also emerges from our analysis regards the effect of a non-vanishing spatial curvature in models with two interacting fluids. Indeed taking the particular case where $\gamma_X=2/3$ for the non-interacting perfect fluid from eq.~(\ref{dotrhoX_if1}) we see that this is equivalent to having a term $\rho\propto a^{-2}$ in the Friedmann equation. Conversely, it is easy to verify that the usual curvature term  taken as $\rho_k =-3k/a^2$ satisfies  
Eqs.~(\ref{Fried_if1}), (\ref{dotH_if1}), (\ref{dotrhoX_if1}) with $\gamma=2/3$. Thus the case of two coupled fluids in $k=-1$ models falls within the scope of our study. 
 
From the definition of $\epsilon$ we have then 
$\epsilon=(2-3\gamma_c)/(3\gamma_d)$ and so, according to our results of Section III,  curvature scaling solutions, as defined in~\cite{Hoogen+Coley+Wands 99}, exist when $\gamma_2<2/3<\gamma_1$. In this case the scale factor evolves as $a\propto t$ and both $\rho_1$ and $\rho_2$ self-adjust to the $\rho_1,\rho_2 \propto a^{-2}$ behaviour of the curvature term. If both $\gamma_1$ and $\gamma_2$ are greater than $2/3$ the origin is a global attractor and we recover the usual asymptotic behaviour found when we have two non-interacting perfect fluids. The curvature term eventually dominates and we have the vanishing of $\rho_1$ and of $\rho_2$. This solution corresponds to well-known Milne universe. 

In the limit case where $\epsilon = -1$, and hence $\gamma_2=2/3$, the curvature term dominates in the future, but only a limited set of solutions corresponds to the depletion of both the $\rho_1$ and the $\rho_2$ components. Indeed almost all solutions end up in the $x=-y$ border line and they correspond to solutions with 
$a\propto t$ which occur with the depletion of the $\rho_1$ component. 
In this case there are no inflationary solutions in the sense that the scale factor does not evolve with a power greater than 1 (we are in the limit coasting model).   
In the alternative limit case, when $\epsilon =1$ and hence $\gamma_1=2/3$, the curvature was the dominating component in the past and the cosmological models evolve either towards scaling solutions characterized by the depletion of the $\rho_2$ component or towards scaling solutions where the curvature vanishes. In this case all the solutions are inflationary.

\subsection{Nucleosynthesis} 
 
It was pointed out in previous works on scaling solutions~\cite{Wetterich 88,WCL 93,Ferreira+Joyce 97,CLW 98} that the most stringent bounds on the admissible densities of the components that are present in addition to the usual perfect fluid are set by the primordial nucleosynthesis of light elements~\cite{Kolb+Turner 90}. 
Since the perfect fluid is radiation, the attractor solution is characterized by the usual of expansion $a\propto t^{1/2}$ and so any deviations from the standard model yields of the light elements are a consequence of the number of degrees of freedom $N(t_{nuc})$ which are due to the extra matter components~\cite{Ferreira+Joyce 97}. The limits that beset this number can be translated into a permitted range of energy density associated to the additional matter contributions which is $ \Omega_{extra} \los 0.13-0.2$~\cite{Ferreira+Joyce 97}. 
   
In the present case $\Omega_{extra}=x$ so that we have the following bounds on $\delta$ at the attractor scaling solution of Eq.~(\ref{3.barx}) 
\begin{equation} 
\delta \gos \frac{(1+\epsilon)^{1-\alpha}\,(1-\epsilon)^{1-\beta}}{0.13^{\alpha+\beta-1}} 
\end{equation} 
when $\alpha+\beta\geq 1$, and  
\begin{equation} 
\delta \los (1+\epsilon)^{1-\alpha}\,(1-\epsilon)^{1-\beta}\, 0.13^{1-\alpha-\beta} 
\end{equation} 
when $\alpha+\beta< 1$. In these expressions $\epsilon=(4-3\gamma_c)/3\gamma_d)$ since $\gamma_X=4/3$. 
 
\subsection{Decay of massive particles} 
We now briefly consider the possibility of applying the present model to a transient regime during the early universe when two particle species interact. For instance the decay of some  massive particle species into a lighter one occurring out of equilibrium. This question has been analysed in the literature (for a review see Chapter 5 of~\cite{Kolb+Turner 90} and references therein) and usually it is assumed that the massive particles  decay into relativistic particles that rapidly thermalize so that, on one hand one may consider them as being a part of the radiation component, and on the other hand one does not have to consider reverse processes.  
 
The model envisaged in this work enables one to relax the assumption of thermal equilibrium of the lighter species. Following~\cite{Kolb+Turner 90} if we denote by $\psi$ the decaying massive  particle species, the relevant equations which are usually adopted are 
\begin{eqnarray} 
\dot{\rho}_\psi+3H\rho_\psi &=& -\tau\,\rho_\psi \label{mpdecay_1}\\ 
\dot{\rho}_{r}+4H\rho_r &=& \tau\,\rho_\psi \label{mpdecay_2} 
\end{eqnarray} 
where $\rho_\psi$ is the energy density of the $\psi$ species, $\rho_r$ is the energy density of the radiation fluid which includes the thermalized daughter products of the decays, and the time at which the decays take place is given by $\tau\sim t\sim H^{-1}$.  
 
If we apply our model it becomes possible to consider the transient stage during which the lighter particle species energy density  increases due to the decays of the $\psi$'s and is not yet in thermal equilibrium with the radiation. Since there is no well-established thermo-kinetic prescription for this situation the purpose of this analysis is mainly illustrative and a more detailed investigation of the specifics of the process is left to a future work. From the viewpoint of our model the usual treatment found in the literature may be associated with taking $\rho_X$ as being the radiation perfect fluid, $\rho_\psi$ to be characterized by $\gamma_2=1$ and the daughter products to be described by $\rho_1$ with $\gamma_1=4/3$.  Then we take $\tau \sim \eta H^{\lambda+1}\,\rho_1^\alpha \rho_2^{\beta-1}$ in Eq.~(\ref{mpdecay_2}) where the parameters are left free. Thus we have a model characterized by $\epsilon=1$ and we immediately know that the dynamics corresponds to the phase-plane of the Fig.~3~c) where only the trajectories above the separatrix that connects the $(0,0)$ singular point to the saddle scaling solution on the $x=1$ border of the phase diagram evolve towards equilibrium with the radiation component. These solutions correspond to a depletion of the $\rho_1$ component and, hence, of the decaying massive particles.  
 
If we relax the assumption that the daughter products are in equilibrium with radiation and we keep the $\gamma_1$ parameter free, we may still have a non-trivial scaling solution in the interior of the phase-plane domain, for which both of the 
components interacting through the decays swiftly adjust themselves to track the perfect fluid behaviour of radiation.  We would have then their thermalization and there is no depletion of the massive particle species in this case. For this to happen the lighter species must be characterized by a $\gamma_1> 4/3$ in order to satisfy the condition $|\epsilon|< 1$.  
According with Eq.~(\ref{3.barx}) the combined density parameter settles at a value determined by $\gamma_1$, since in the present case it completely defines $\epsilon$ (we have $\epsilon=(5-3\gamma_1)/(3(\gamma_1-1))$, and also by the values taken for $\alpha$, $\beta$ and $\delta$ (the latter is given by $\delta=3(3H/2)^{\alpha+\beta-2} \rho_1^{-1}\rho_2^{1-\beta}\tau^{-1}$). However as presented in our study additional conditions must also be met in order to have an attracting NTSS. The relevant cases are those represented in Figs.~4 c), 5~c) and d) and 7 a) and a common condition $\delta > (1+\epsilon)^{1-\alpha}(1-\epsilon)^{1-\beta}$ which translates into  
\begin{equation} 
\delta > \left(\frac{2}{3}\right)^{\alpha+\beta-2} \, \left(\frac{1}{\gamma_1-1}\right)^{1-\alpha}\,  
 \left(\frac{4-3\gamma_1}{\gamma_1-1}\right)^{1-\beta} \; . 
\end{equation} 
If one selects particular values of the remaining parameters this means yet another  constraint on $\delta$.  
  
The main point to emphasize is then that there exists the possibility within the framework of a flat model of having a dynamical thermalization of the three components and, hence, the energy densities of the particle species adding to the radiation energy density. In the conventional models the relics of the $\psi$'s do not contribute to the latter and are subject to the so-called Lee-Weinberg bound~\cite{Kolb+Turner 90}.

\subsection{Conclusion} 
To conclude we want to stress that our results reveal how  the consideration of nonlinear interactions between some  of the matter components of the universe allows for the emergence of a variety of phenomena of which the  scaling solutions and oscillatory behaviour are remarkable examples. In this sense, our study resembles the approach of Ref.~\cite{B+K 77} in their analysis of the dynamics of models with bulk viscosity. Apart from  
producing the classification of the models in parameter space in terms of the qualitative behaviour of the solutions, we have also addressed some of the cases that can be singled out not only from the point of view of their dynamics, but also  
from the perspectives of possible application to the thermal physics of the universe. From this more physical standing, we would like to stress the following aspects. First, our models generalize the scalar field models yielding scaling behaviour thus providing a different physical setting in which the interesting properties of scaling solutions may be obtained. 
Second, the classification scheme provides the guidelines to be followed in the search for the possible causes of a given phenomenology. Third, the detection of new phenomena or the need to re-evaluate observational limits regarding, for instance, relic abundances, nucleosynthesis, dark matter, may be reexamined with the help of the classification presented here. However, needless to say that the consideration of any specific model requires a detailed kinetic analysis to provide a solid justification to the particular interaction model.  
Finally, our study has lead naturally to new results on the existence of curvature scaling solutions (sub-sect. IV-B) and on the conditions for finite inflationary periods (sub-sect. IV-A). 
 


\begin{figure} 
\begin{center} 
\begin{tabular}{lcccccr}
\psfig{file=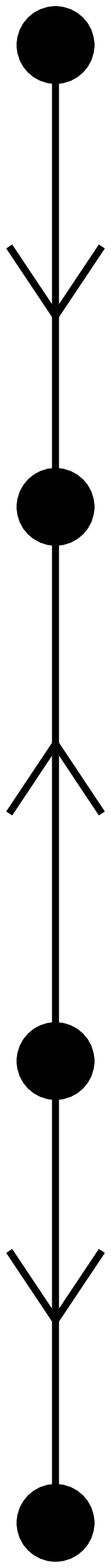,height=80pt,width=5pt} & \ \ & \ \ & 
\psfig{file=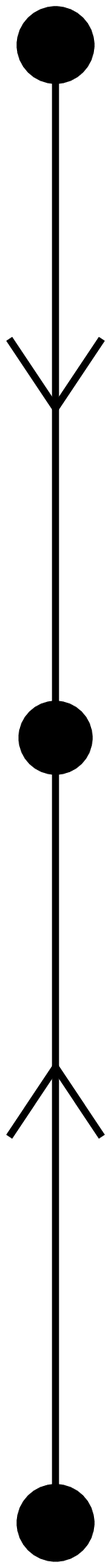,height=80pt,width=5pt} & \  \ & \ \ &
\psfig{file=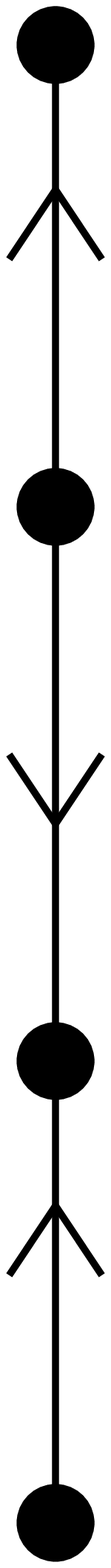,height=80pt,width=5pt} \\
  a) & \ \ & \ \ & b) & \ \ & \ \ & c)
\end{tabular} 
\end{center} 
\caption{Non trivial scaling solutions on the invariant line 
$x=1$.  a) $\alpha ,\beta >1$, $ (\frac {2}{\alpha +\beta -2} 
 )^{\alpha +\beta -2}(\alpha -1)^{\alpha -1} 
(\beta -1)^{\beta -1} > \frac {1}{\delta }$. b) $\alpha <1<\beta 
$ or $\alpha =1$, $\beta >1$ and $2^{\beta -1} 
> \frac {1}{\delta }$, or  $\alpha < 1$, $\beta =1$ and $2^{\alpha -1} 
< \frac {1}{\delta }$. 
c)  $\alpha ,\beta <1$, $ (\frac {2}{2-\alpha -\beta } 
 )^{\alpha +\beta -2}(1-\alpha )^{\alpha -1} 
(1-\beta )^{\beta -1} < \frac {1}{\delta }$. } \label{fig1} 
\end{figure} 


\begin{figure} 
\begin{center} 
\begin{tabular}{ccc}
\hspace*{-0.5cm}
\psfig{file=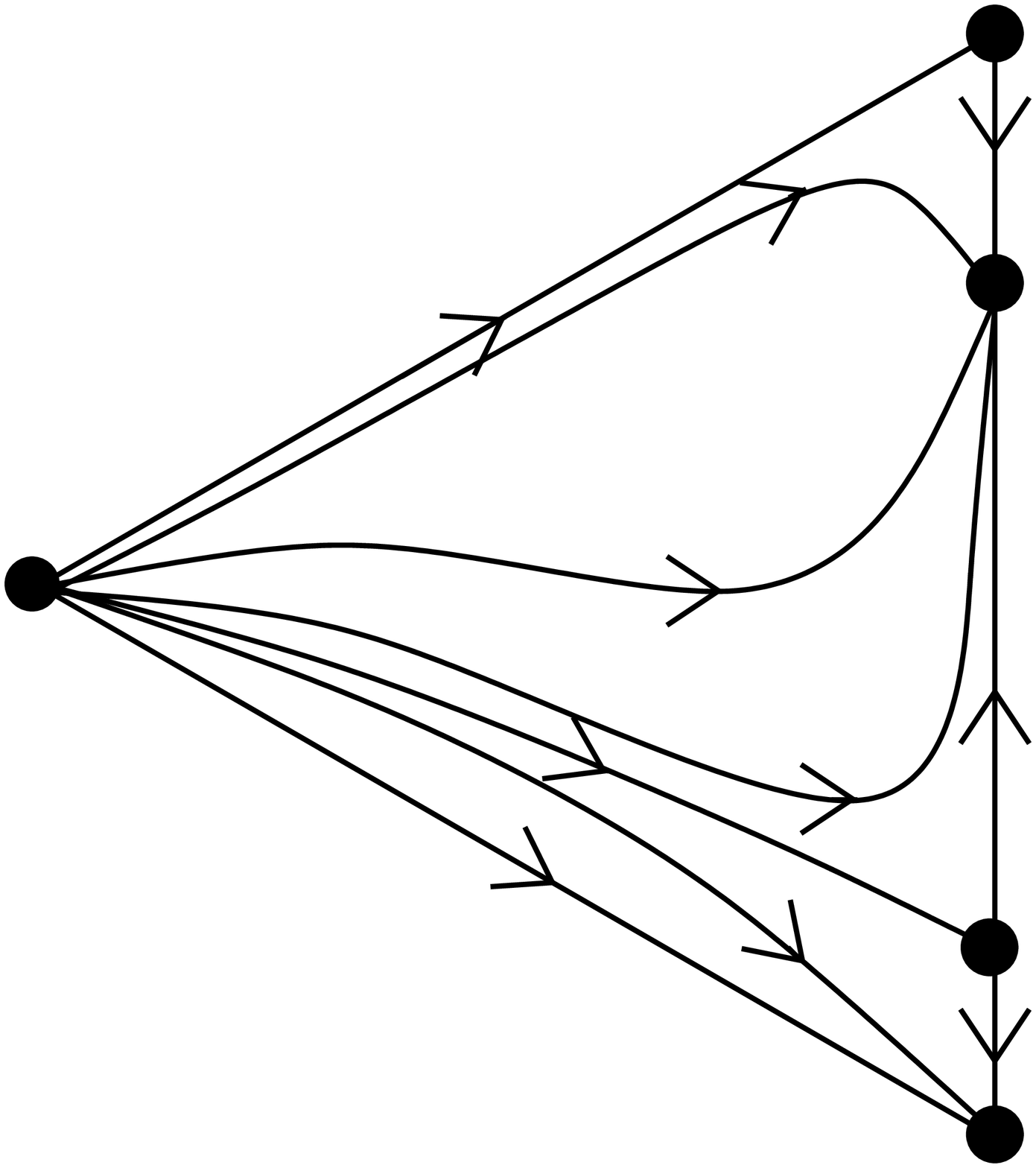,height=80pt,width=80pt}  & 
\hspace*{0.5cm}
\psfig{file=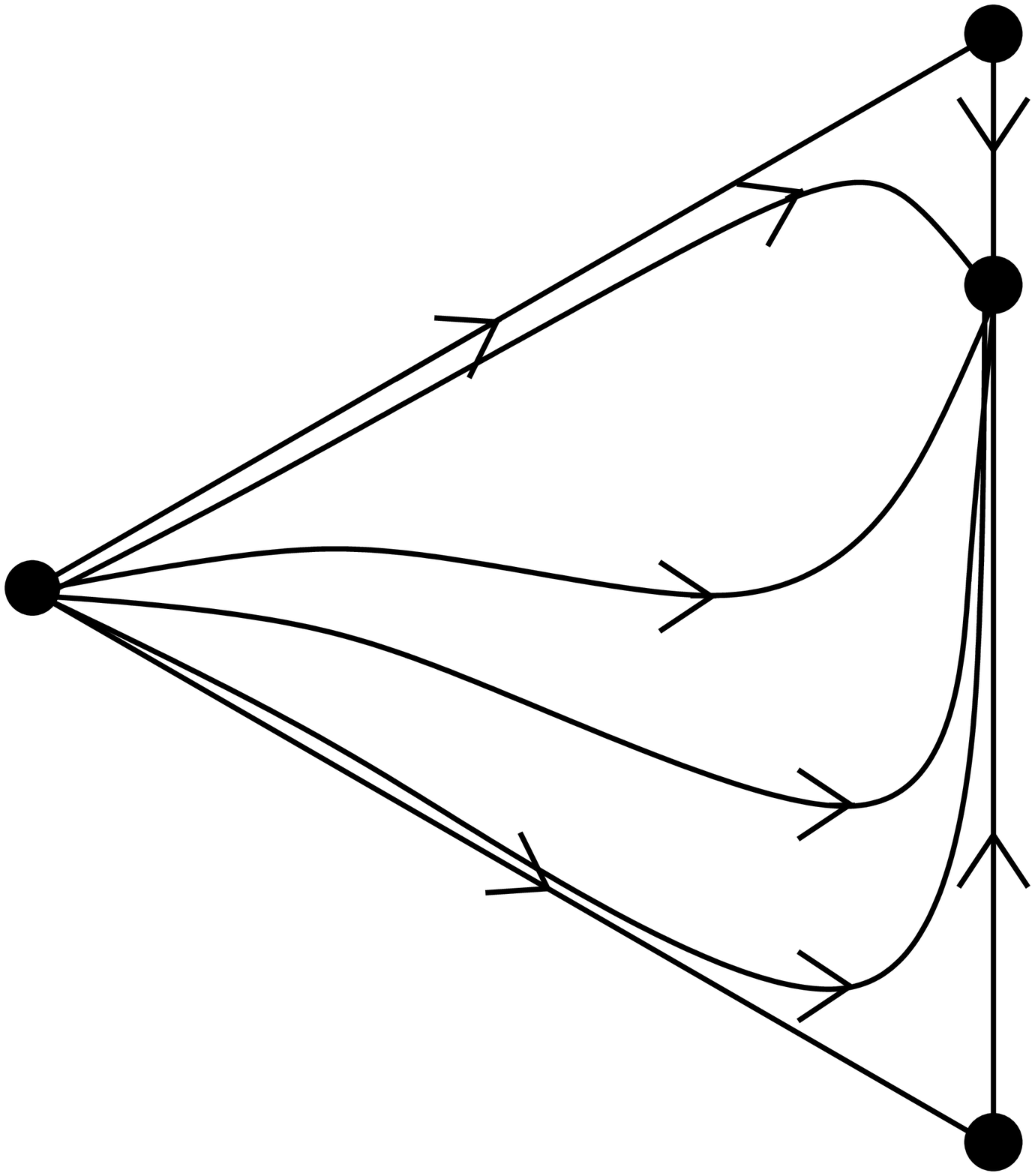,height=80pt,width=80pt}  &  
\hspace*{0.5cm}
\psfig{file=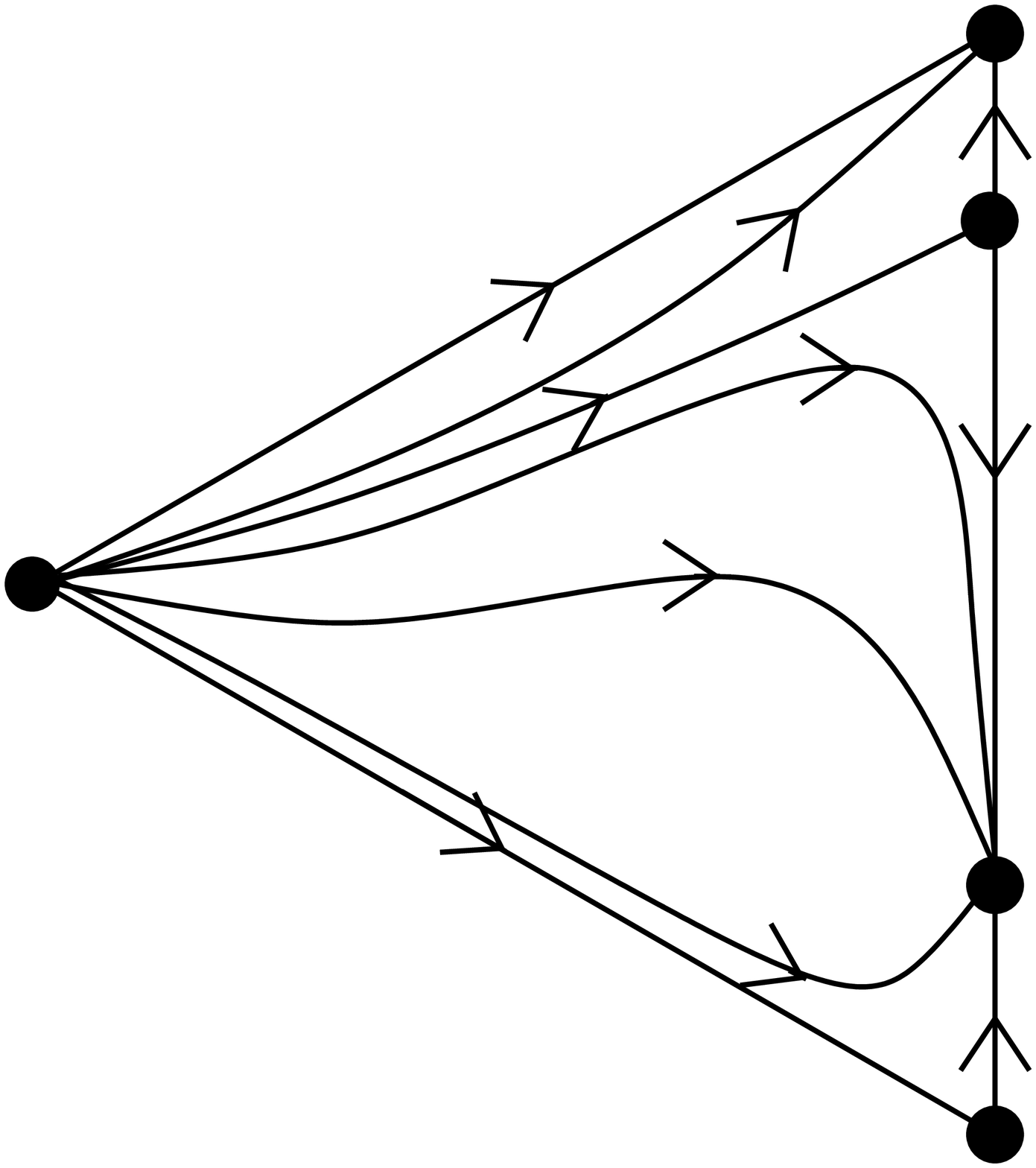,height=80pt,width=80pt} \\ a) & b) & c)
\end{tabular} 
\end{center} 
\caption{Global dynamics for $\epsilon >1$.  a) $\alpha ,\beta 
>1$, $ (\frac {2}{\alpha +\beta -2} )^{\alpha +\beta -2}(\alpha 
-1)^{\alpha -1} (\beta -1)^{\beta -1} > \frac {1}{\delta }$. b) 
$\alpha <1<\beta $ or $\alpha =1$, $\beta >1$ and $2^{\beta -1} 
> \frac {1}{\delta }$, or  $\alpha < 1$, $\beta =1$ and $2^{\alpha -1} 
< \frac {1}{\delta }$. 
c)  $\alpha ,\beta <1$, $ (\frac {2}{2-\alpha -\beta } 
 )^{\alpha +\beta -2}(1-\alpha )^{\alpha -1} 
(1-\beta )^{\beta -1} < \frac {1}{\delta }$. } \label{fig2} 
\end{figure} 

 
\begin{figure} 
\begin{center} 
\begin{tabular}{ccc}
\hspace*{-0.5cm}
\psfig{file=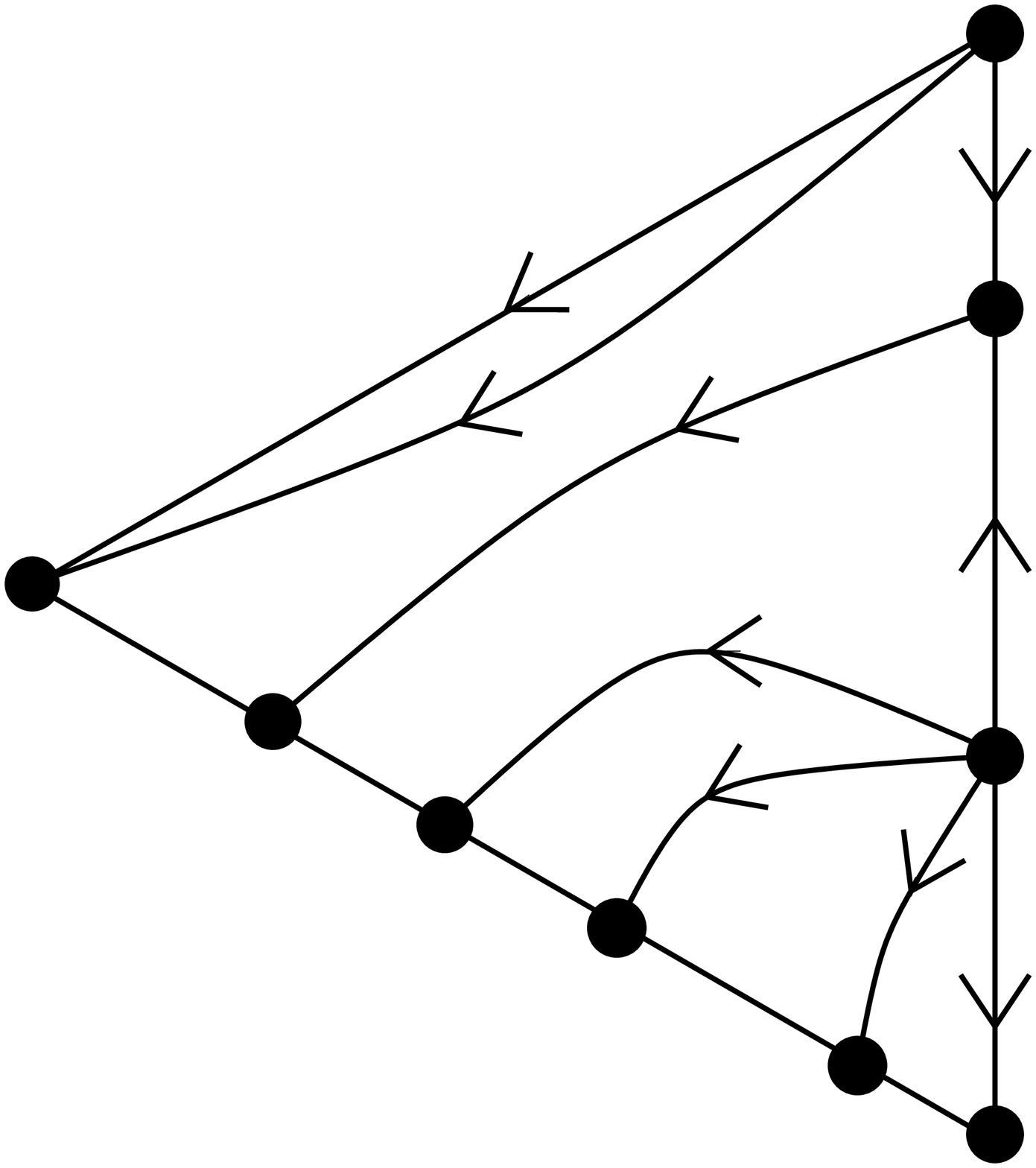,height=80pt,width=80pt}  & 
\hspace*{0.5cm}
\psfig{file=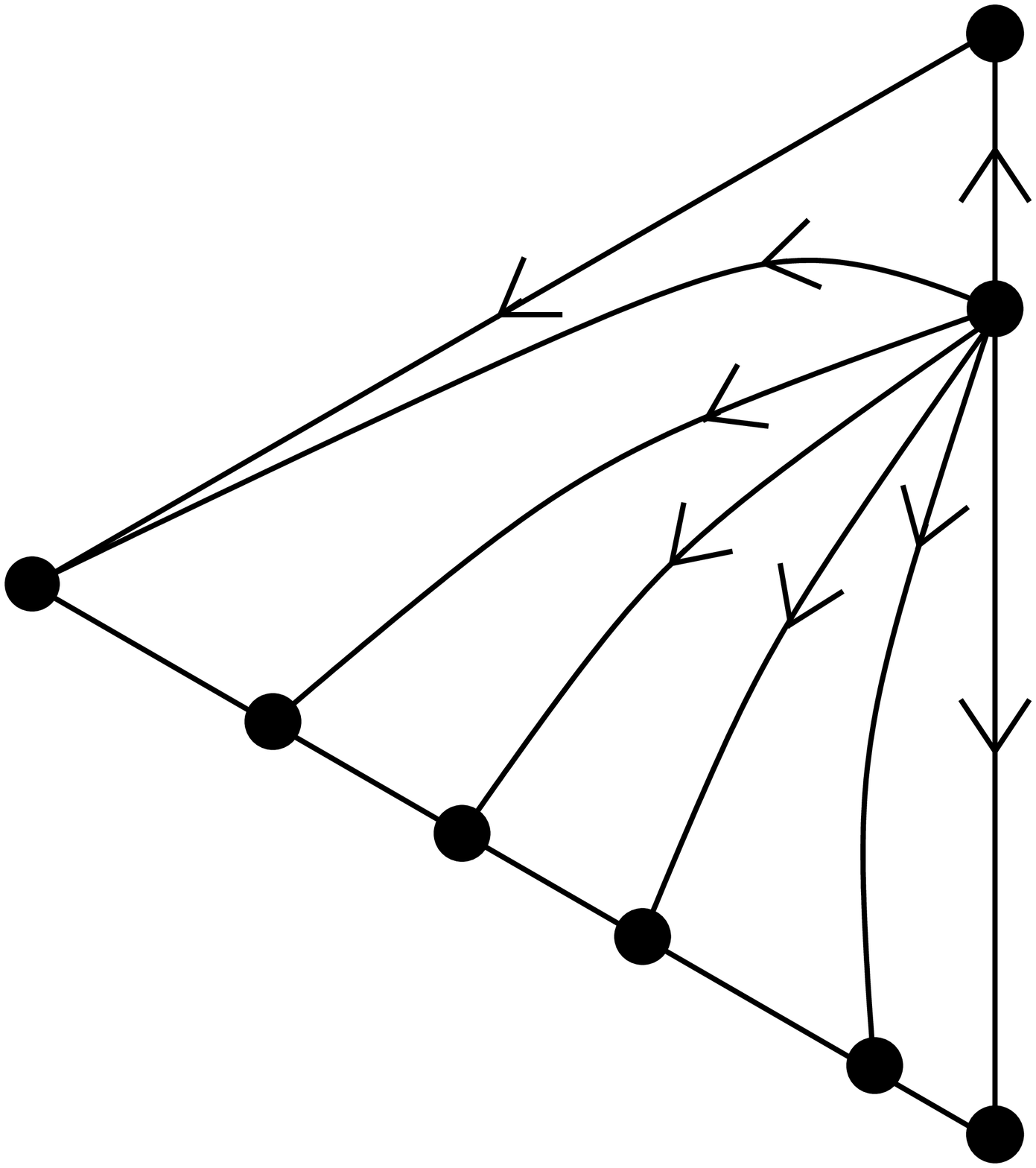,height=80pt,width=80pt}  &  
\hspace*{0.5cm}
\psfig{file=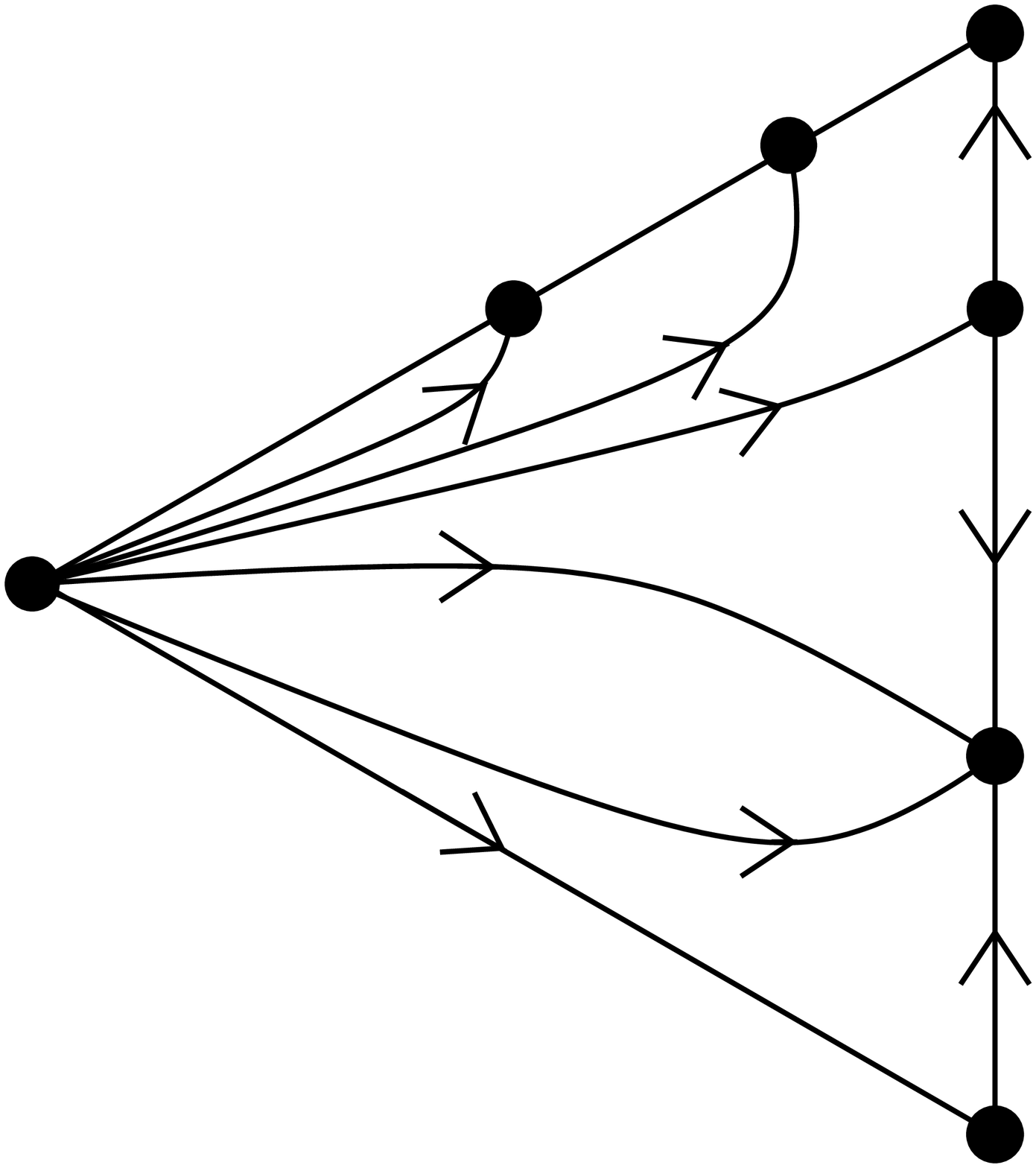,height=80pt,width=80pt} \\ a) & b) & c)
\end{tabular} 
\end{center} 
\caption{Non trivial scaling solutions for $|\epsilon |=1$. 
 a) $\alpha ,\beta $ as in Figure 2.a) and $\epsilon =-1$. 
b) $\epsilon =-1$ and $\beta <1<\alpha $ or $\beta =1$, $\alpha 
>1$ and $2^{\alpha -1} 
> \frac {1}{\delta }$, or  $\beta < 1$, $\alpha =1$ and $2^{\beta -1} 
< \frac {1}{\delta }$. 
c)  $\alpha ,\beta $ as in Figure 2.c) and $\epsilon =1 $. } 
\label{fig3} 
\end{figure} 


\begin{figure} 
\begin{center} 
\begin{tabular}{ccc}
\hspace*{-0.5cm}
\psfig{file=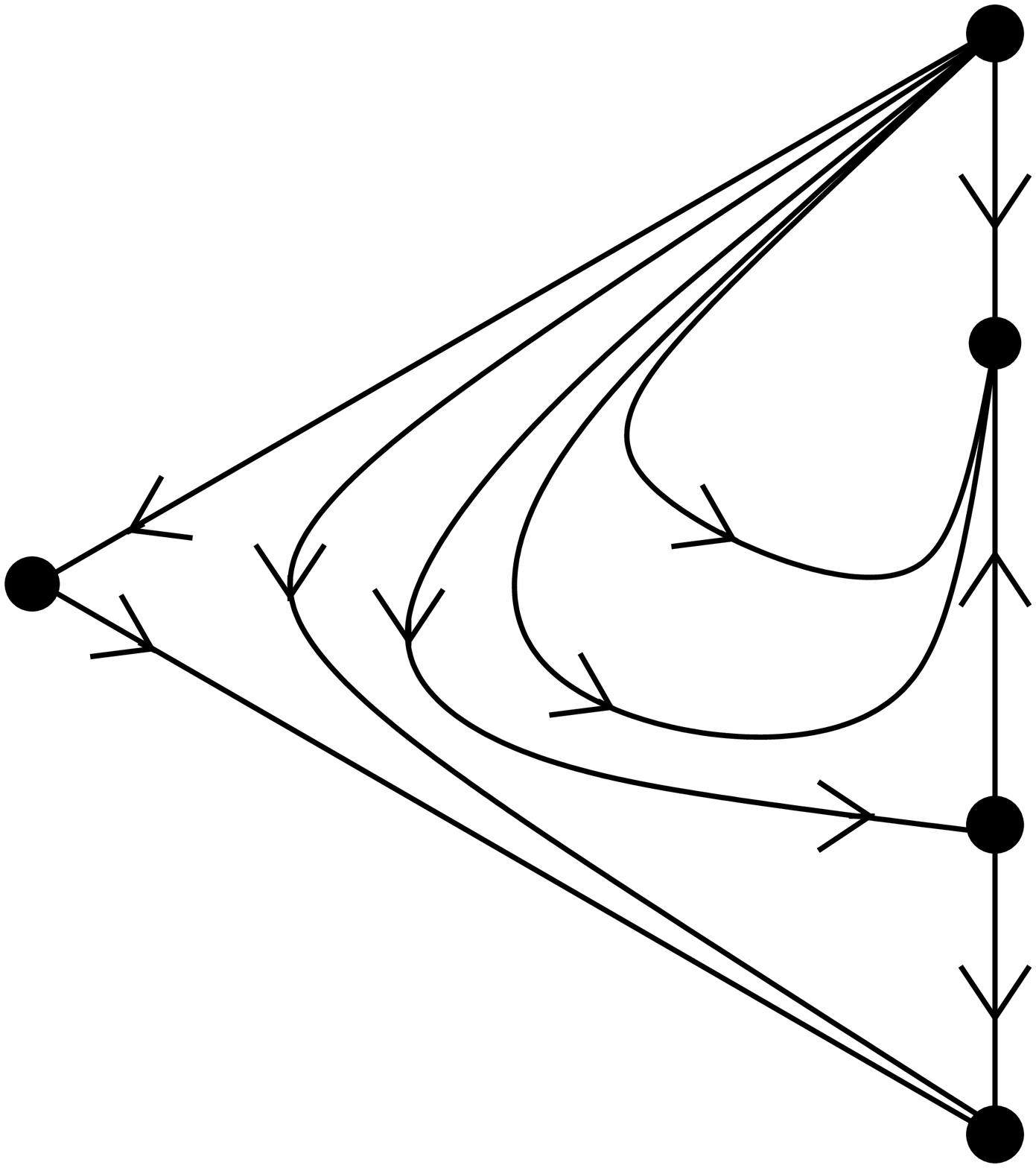,height=80pt,width=80pt}  & 
\hspace*{0.5cm}
\psfig{file=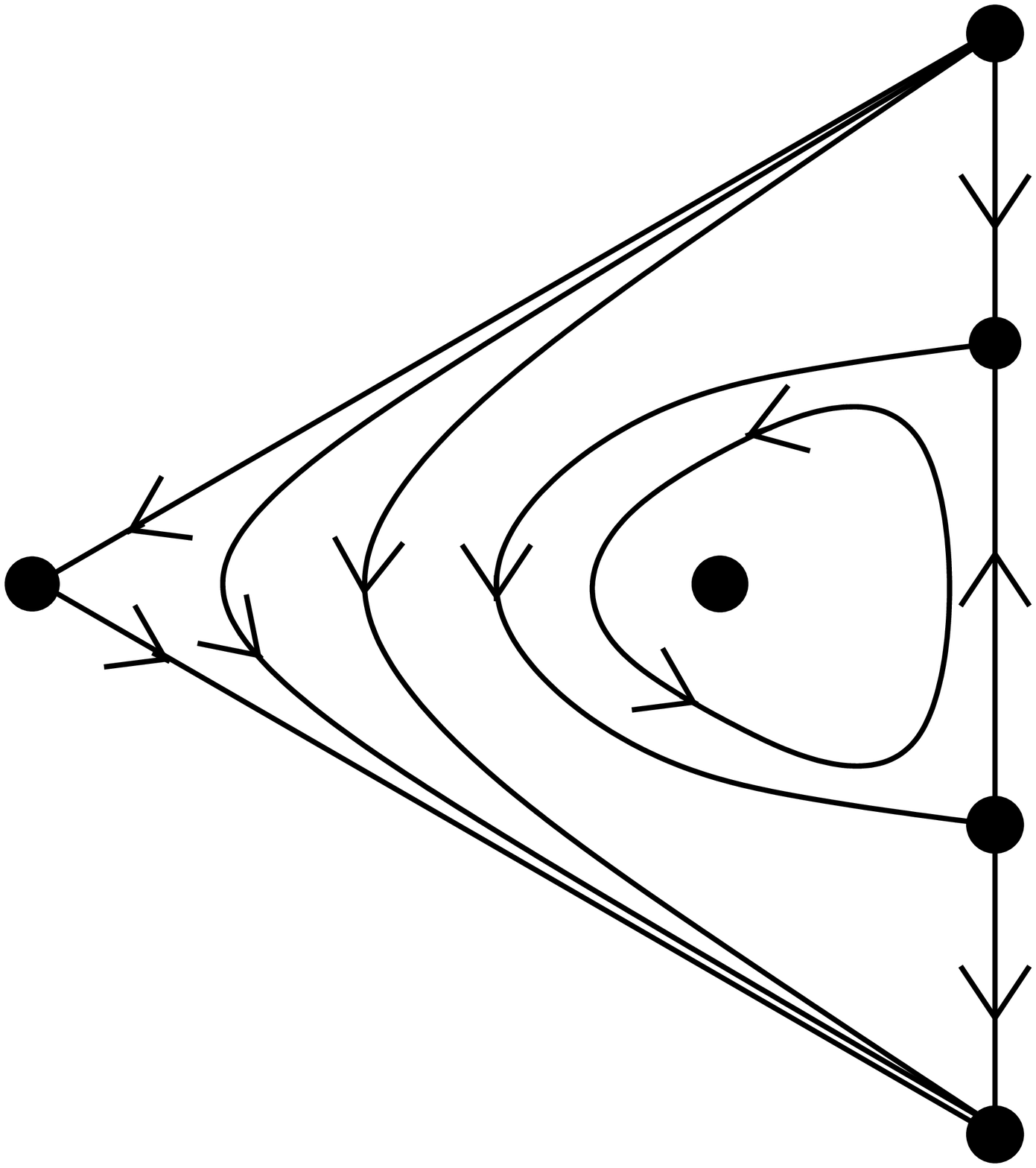,height=80pt,width=80pt}  &  
\hspace*{0.5cm}
\psfig{file=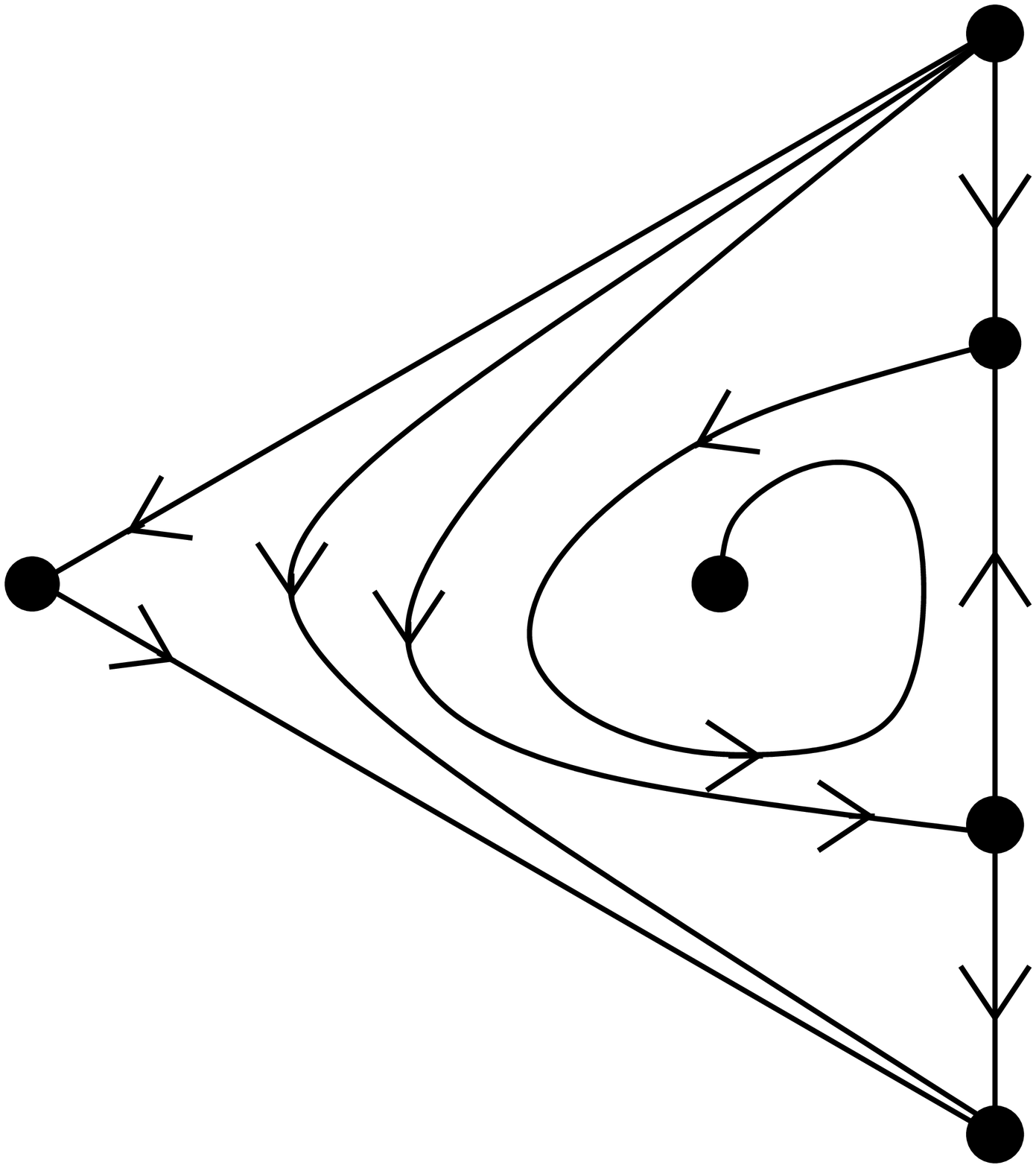,height=80pt,width=80pt} \\ a) & b) & c)
\end{tabular} 
\end{center} 
\caption{Non trivial scaling solutions for $|\epsilon |<1$ and 
$\alpha ,\beta >1$. 
 a) $(1+\epsilon )^{\alpha -1}(1-\epsilon )^{\beta -1} \leq 1/\delta $ 
 and $ (\frac {2}{\alpha +\beta -2} 
 )^{\alpha +\beta -2}(\alpha -1)^{\alpha -1} 
(\beta -1)^{\beta -1} > 1/\delta $. b)  $(1+\epsilon )^{\alpha 
-1}(1-\epsilon )^{\beta -1} > 1/\delta $ and $\epsilon = \epsilon 
_M$. c)  $(1+\epsilon )^{\alpha -1}(1-\epsilon )^{\beta -1} > 
1/\delta $ and $\epsilon > \epsilon _M$. } \label{fig4} 
\end{figure} 


\begin{figure} 
\begin{center} 
\begin{tabular}{ccc}
\hspace*{-0.5cm}
\psfig{file=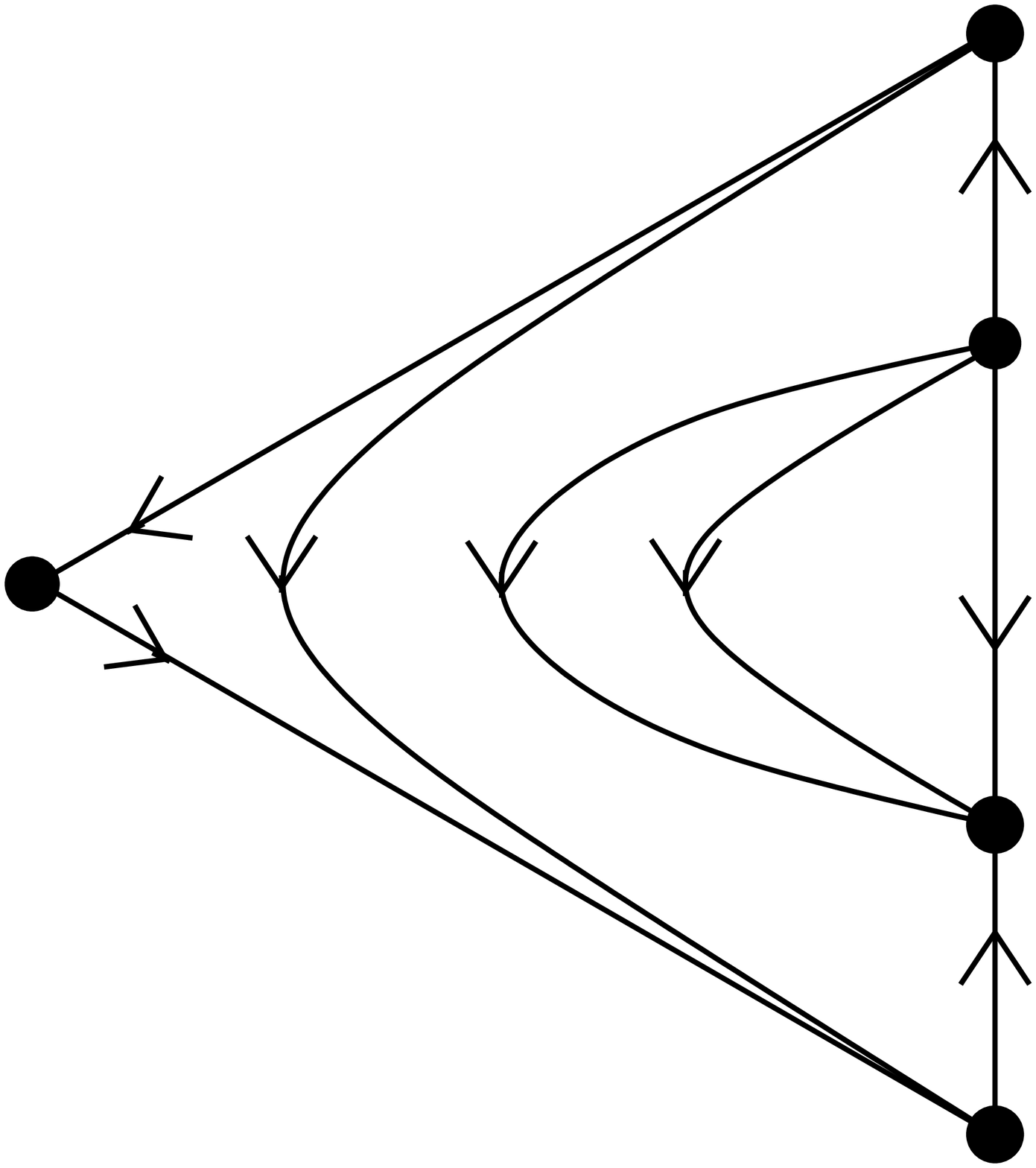,height=80pt,width=80pt}  & 
\hspace*{0.5cm}
\psfig{file=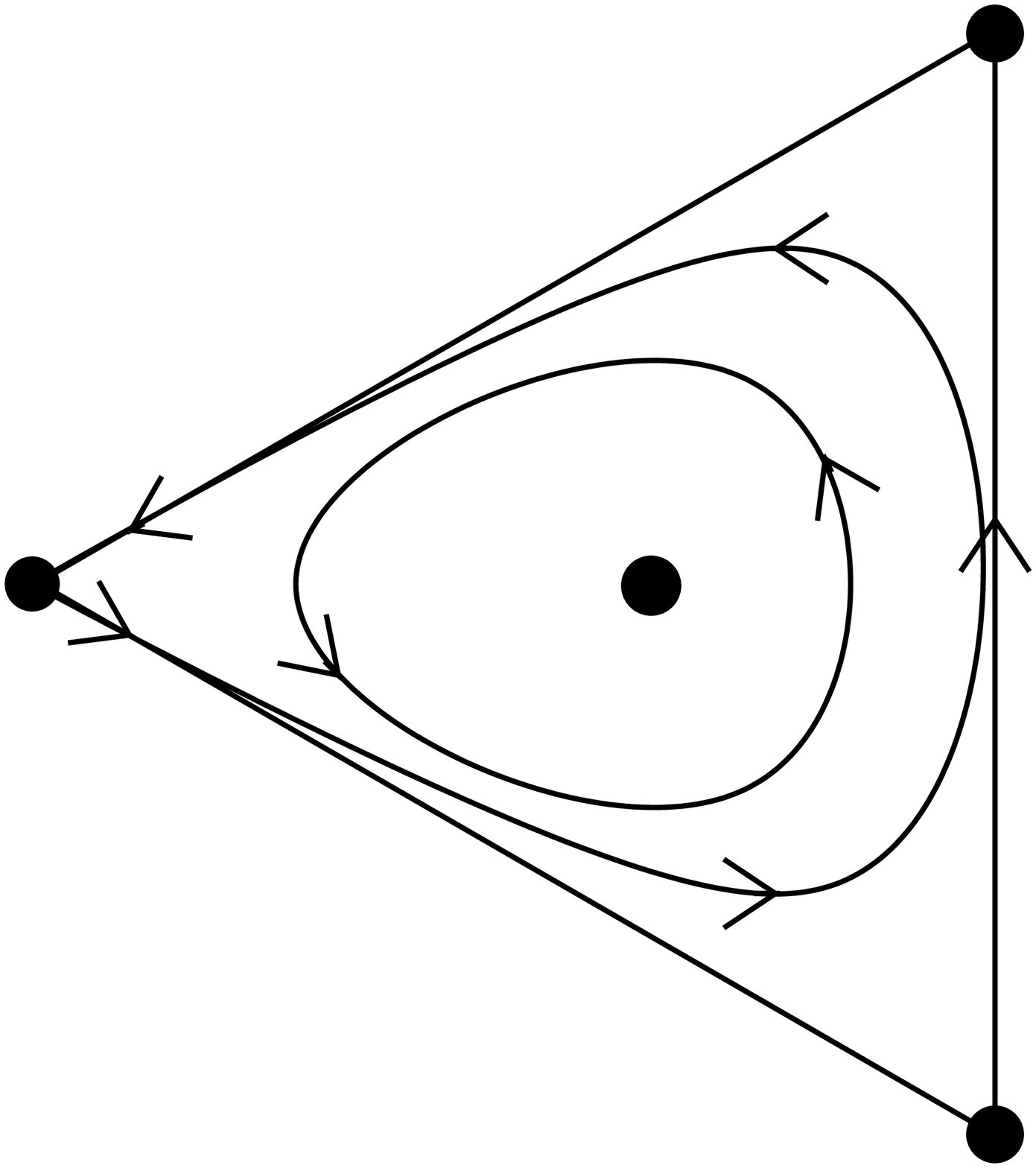,height=80pt,width=80pt}  &  
\hspace*{0.5cm}
\psfig{file=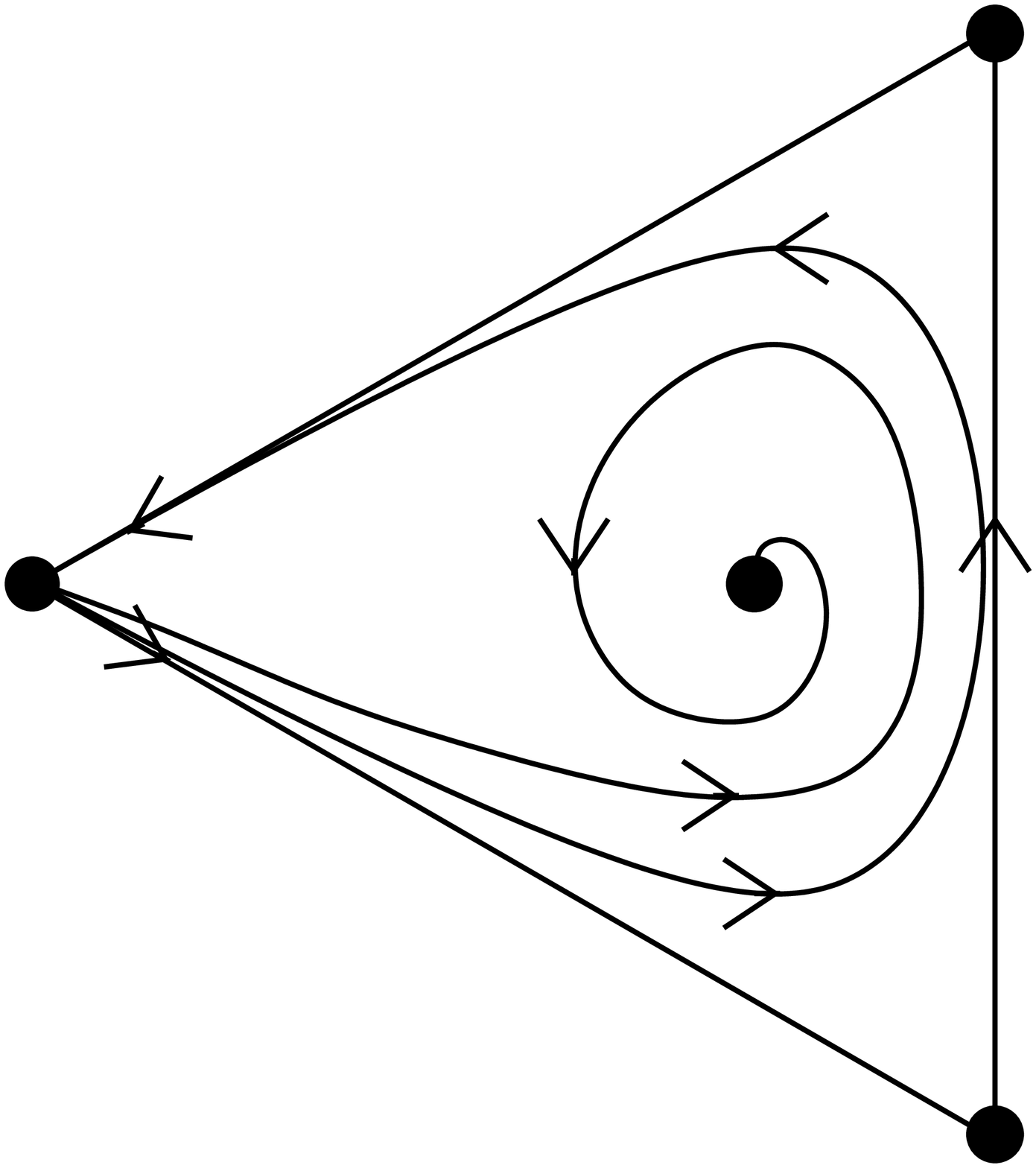,height=80pt,width=80pt} \\ a) & b) & c)
\end{tabular}
\end{center}
\vskip 1cm
\begin{center}
\begin{tabular}{cc}
\hspace*{-0.5cm}
\psfig{file=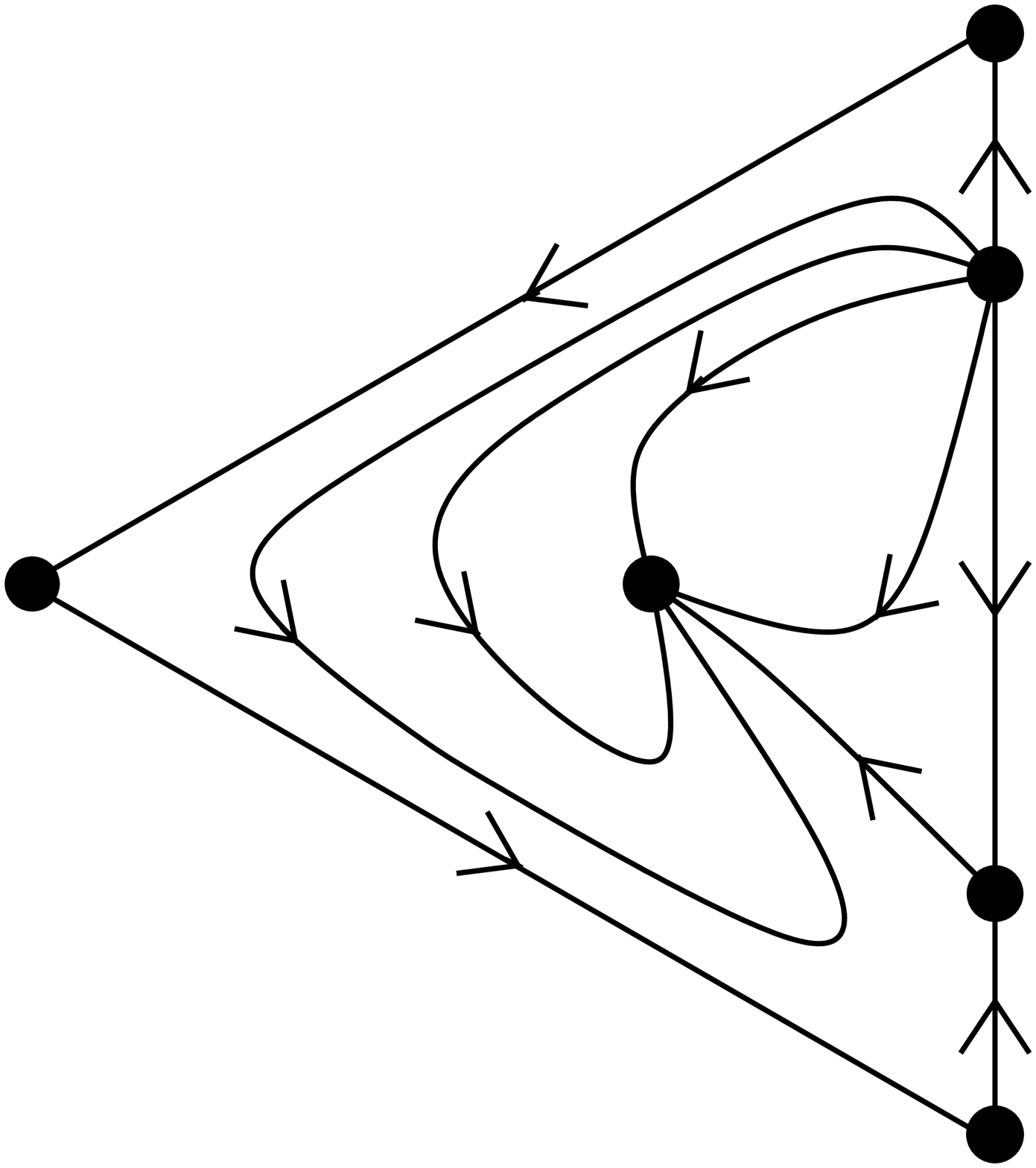,height=80pt,width=80pt}  & 
\hspace*{0.5cm}
\psfig{file=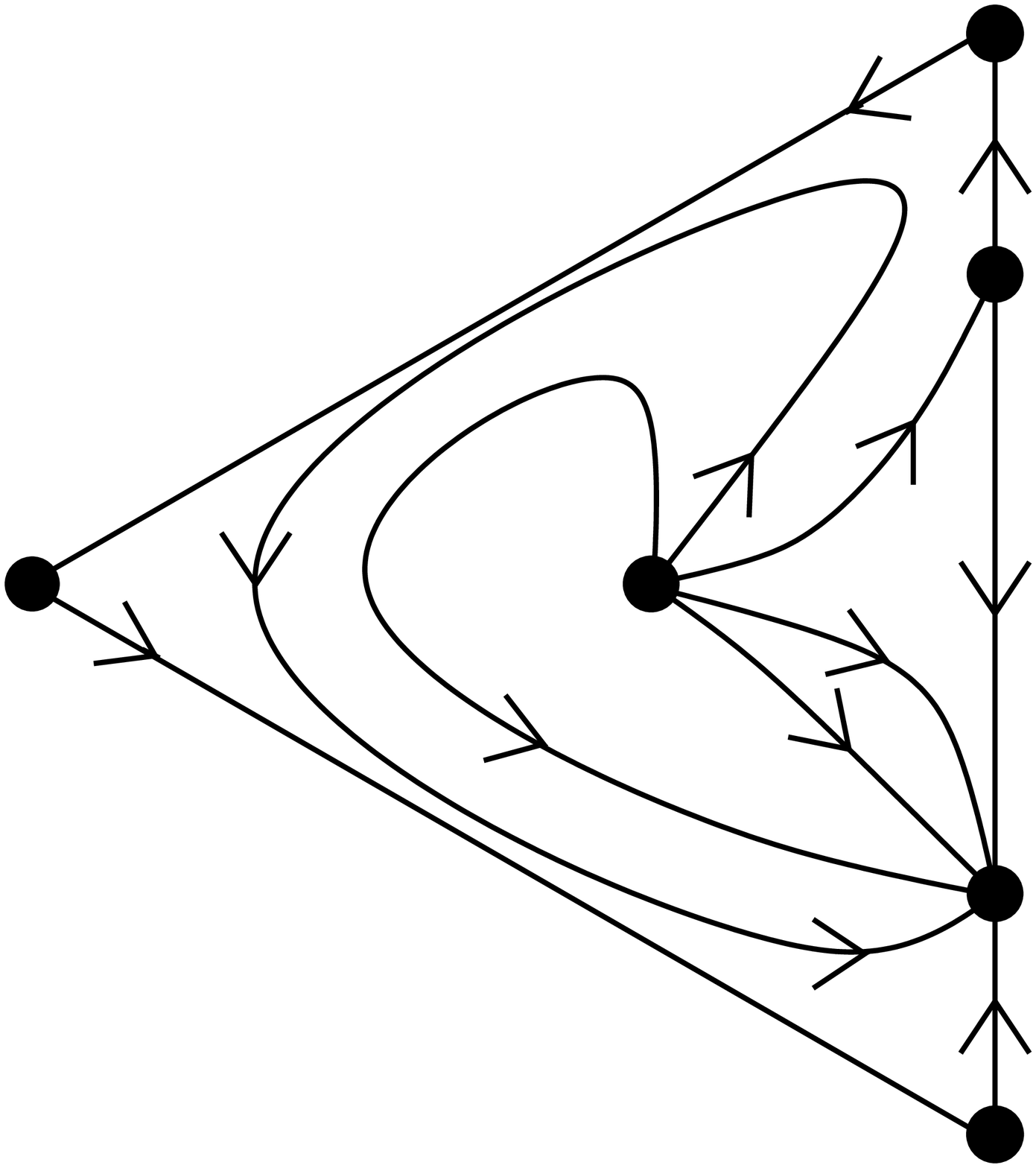,height=80pt,width=80pt}   \\ d) & e)
\end{tabular} 
\end{center} 
\caption{Non trivial scaling solutions for $|\epsilon |<1$, 
$\alpha ,\beta <1$, and $\alpha + \beta \geq 1$}. a)$(1+\epsilon 
)^{1-\alpha }(1-\epsilon )^{1-\beta } \geq \delta$. b),c) 
$(1+\epsilon )^{1-\alpha }(1-\epsilon )^{1-\beta } < \delta 
 $ and $ (\frac {2}{2-\alpha -\beta } 
 )^{2-\alpha -\beta }(1-\alpha )^{1-\alpha } 
(1-\beta )^{1-\beta } \leq \delta $. 
 b) $\epsilon =\epsilon _M$. c) $\epsilon <\epsilon _M$. 
d), e) $(1+\epsilon )^{1-\alpha }(1-\epsilon )^{1-\beta } < \delta 
 $ and $ (\frac {2}{2-\alpha -\beta } 
 )^{2-\alpha -\beta }(1-\alpha )^{1-\alpha } 
(1-\beta )^{1-\beta } > \delta $. d) $\epsilon <\epsilon _M$. e) 
$\epsilon >\epsilon _M$. 
 \label{fig5} 
\end{figure} 


\begin{figure} 
\begin{center} 
\begin{tabular}{cc}
\hspace*{-0.5cm}
\psfig{file=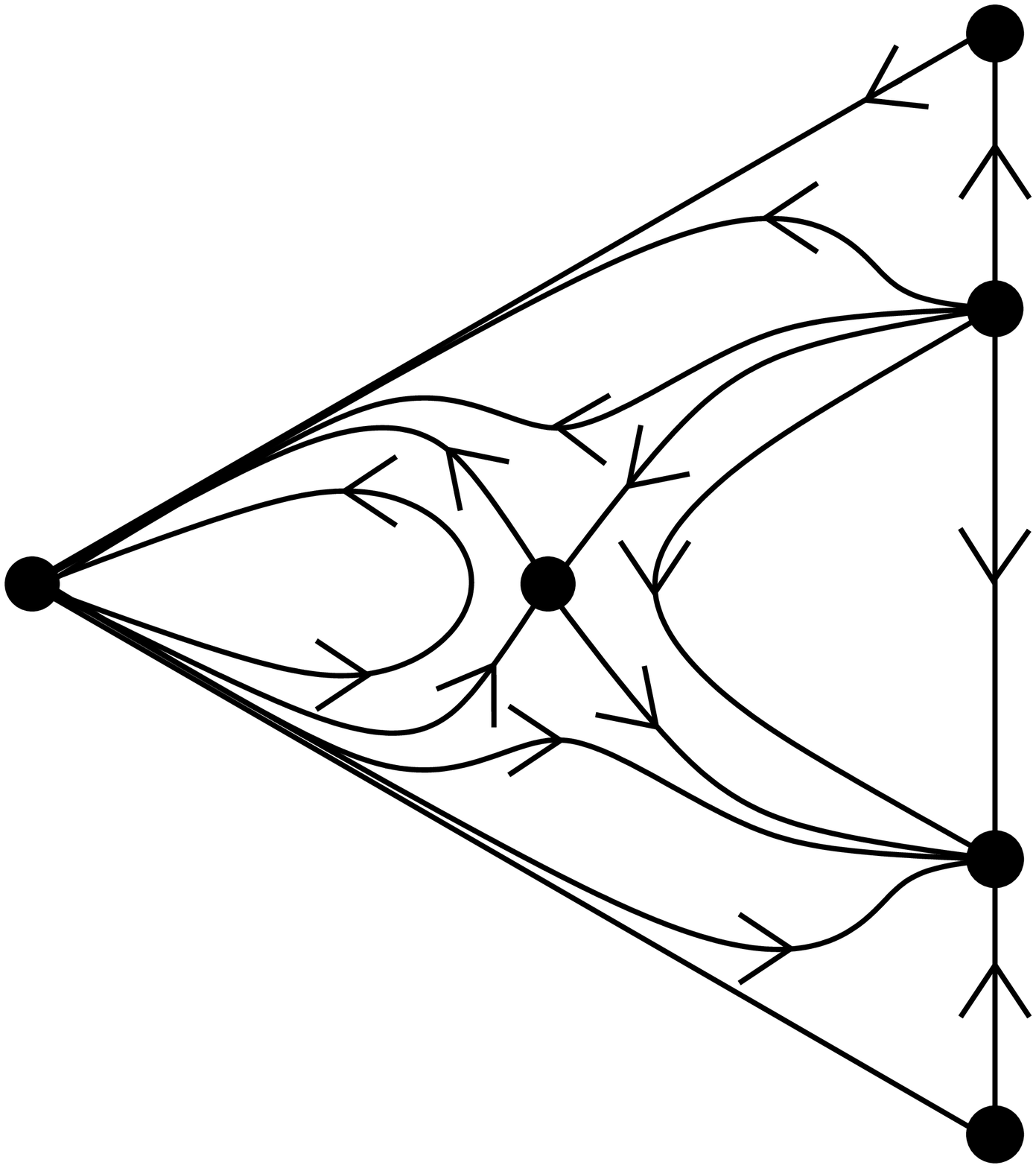,height=80pt,width=80pt}  & 
\hspace*{0.5cm}
\psfig{file=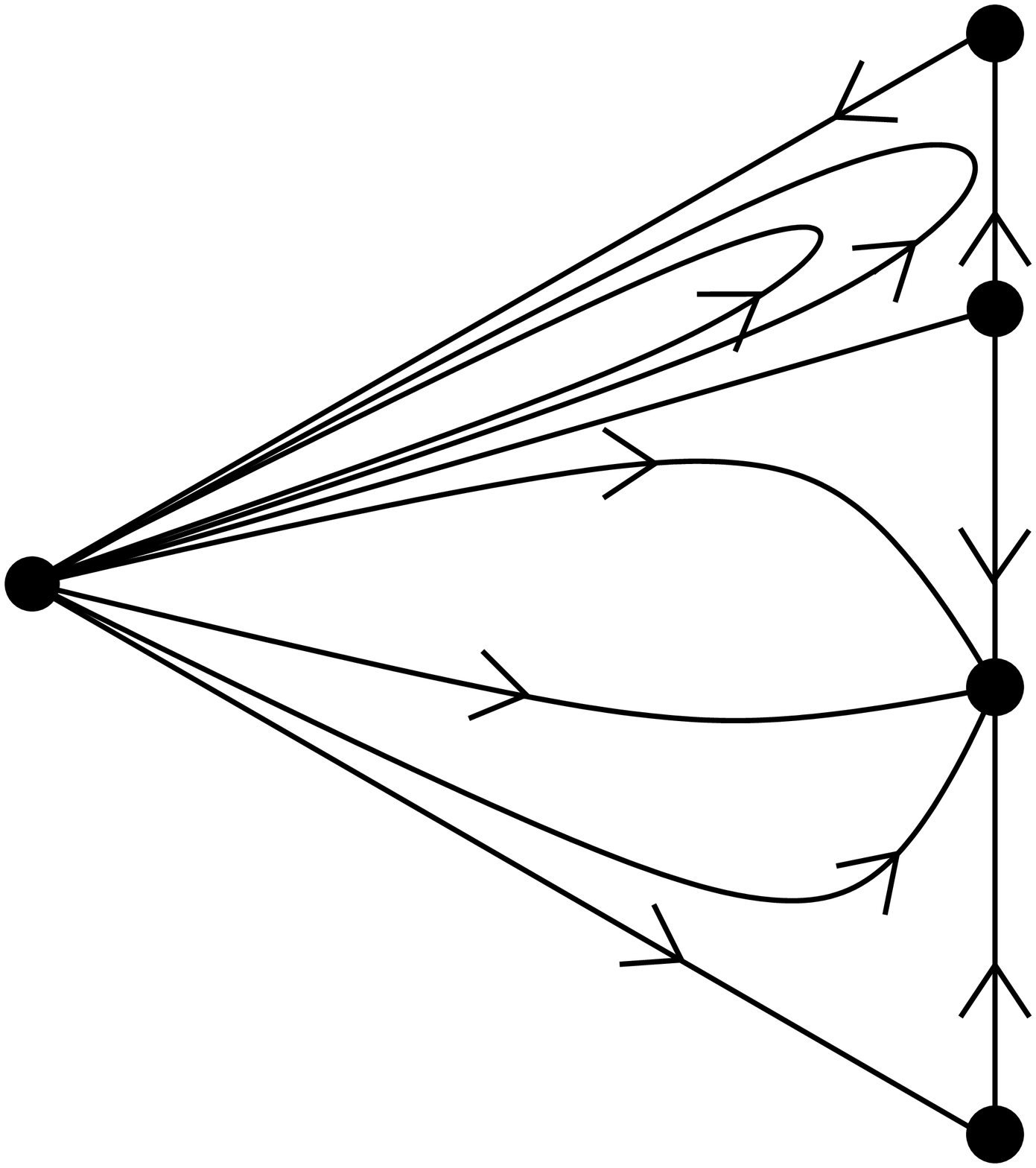,height=80pt,width=80pt} \\ a) & b) 
\end{tabular} 
\end{center} 
\caption{Non trivial scaling solutions for $|\epsilon |<1$, 
$\alpha ,\beta <1$, and $\alpha + \beta < 1$}. a)$(1+\epsilon 
)^{1-\alpha }(1-\epsilon )^{1-\beta } > \delta$. b) $(1+\epsilon 
)^{1-\alpha }(1-\epsilon )^{1-\beta } \leq \delta 
 $, $\epsilon >0$  and $ (\frac {2}{2-\alpha -\beta } 
 )^{2-\alpha -\beta }(1-\alpha )^{1-\alpha } 
(1-\beta )^{1-\beta } > \delta $. 
  \label{fig6} 
\end{figure}


\begin{figure} 
\begin{center} 
\begin{tabular}{cc}
\hspace*{-0.5cm}
\psfig{file=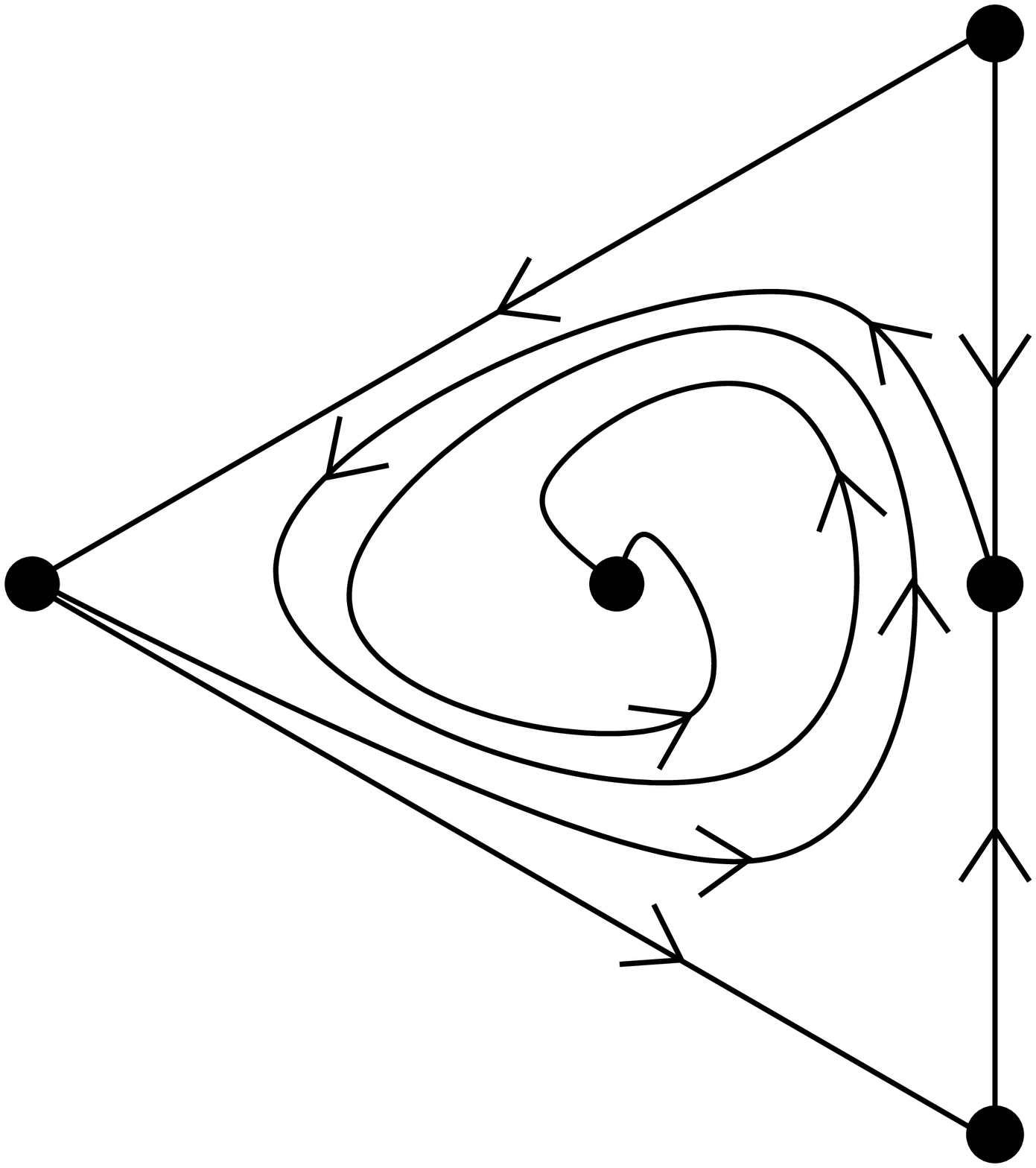,height=80pt,width=80pt}  & 
\hspace*{0.5cm}
\psfig{file=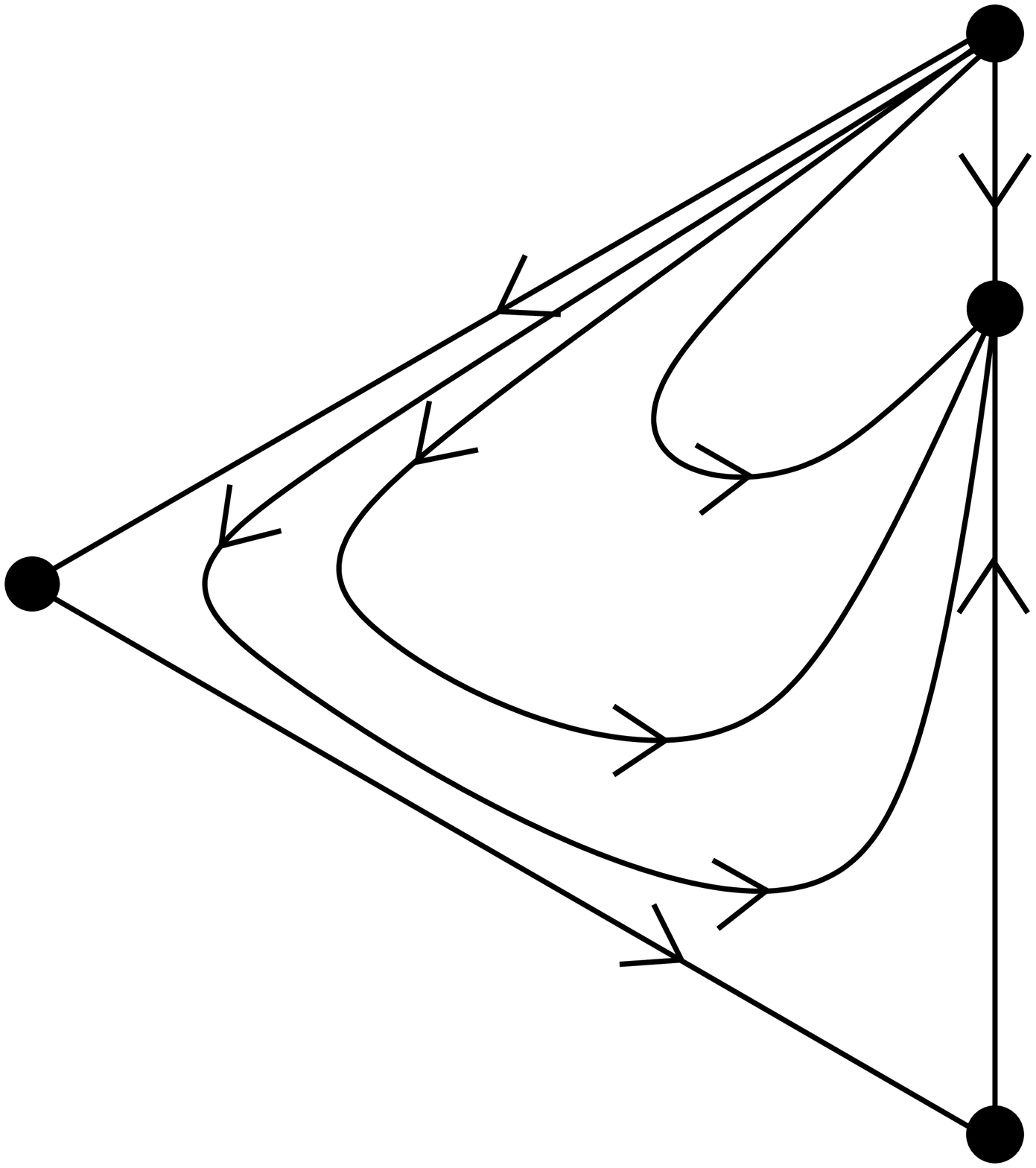,height=80pt,width=80pt} \\ a) & b) 
\end{tabular} 
\end{center} 
\caption{Non trivial scaling solutions for $|\epsilon |<1$ and 
$\alpha \leq 1 < \beta $ or $\alpha < 1 \leq \beta $. 
a)$(1+\epsilon )^{\alpha -1}(1-\epsilon )^{\beta -1} 
> 1/\delta$. b) $(1+\epsilon )^{\alpha -1}(1-\epsilon )^{\beta -1} 
\leq 1/\delta $, and $\alpha < 1<\beta $ or $\alpha =1$, $\beta 
>1$ and $2^{\beta -1}>1/\delta $, or $\alpha <1$, $\beta 
=1$ and $2^{\alpha -1}<1/\delta $.} 
  \label{fig7} 
\end{figure} 


\begin{figure} 
\center
\psfig{file=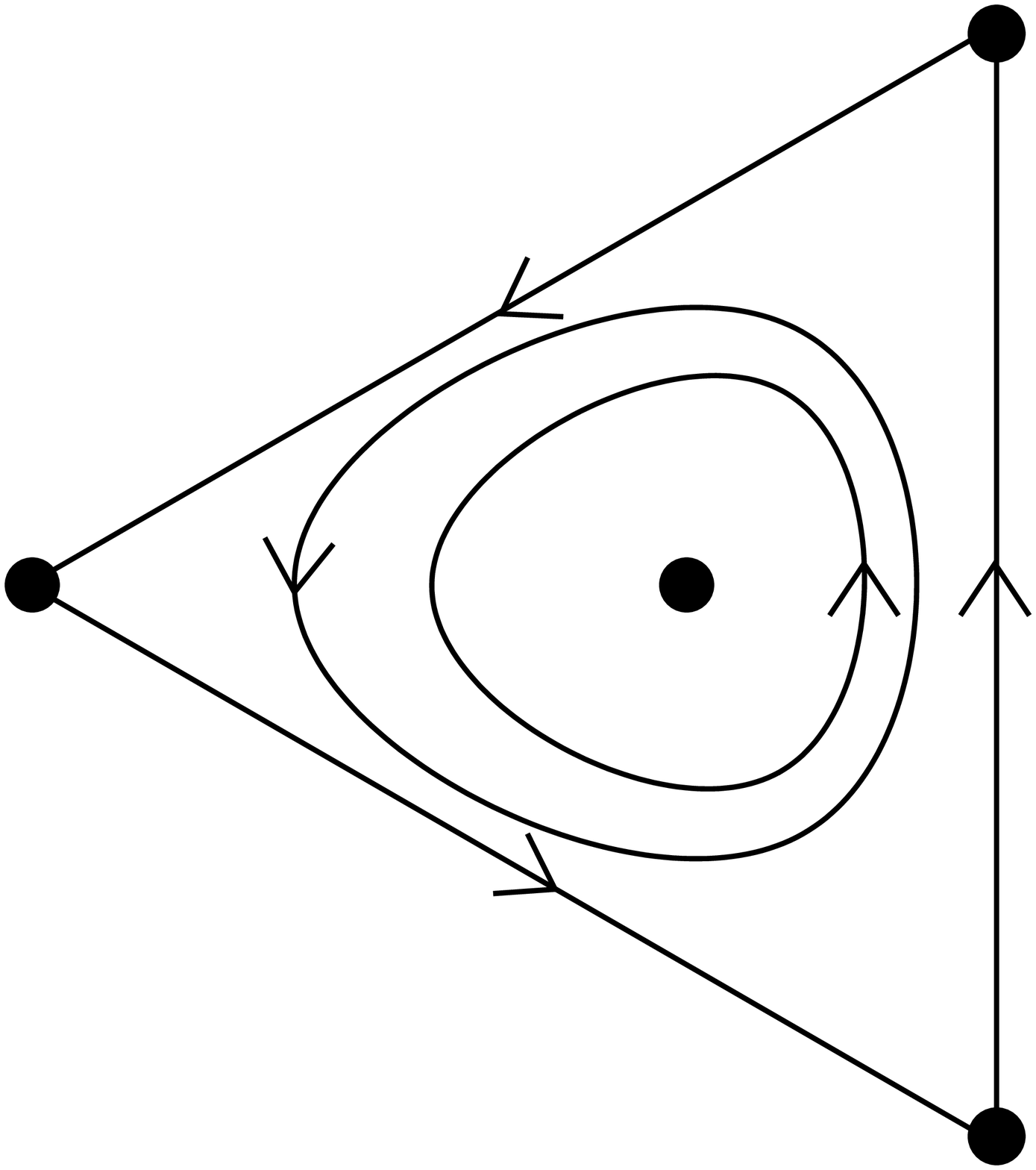,height=80pt,width=80pt}  
\caption{Non trivial scaling solutions for $|\epsilon |<1$ and 
$\alpha = \beta =1 $ and $\delta >1$.} 
  \label{fig8} 
\end{figure} 


\begin{thebibliography}{00} 
\bibitem{Wainright + Ellis 97} J. Wainwright, G Ellis, {\em Dynamical systems in Cosmology}, Cambridge Univ. Press (1997).  
\bibitem{Zlatev+Wang+Steinhardt 99} I. Zlatev, L. Wang and P.J. Steinhardt Phys. Rev. Lett. {\bf 82}, 896 (1999). 
\bibitem{Ratra+Peebles 88} B. Ratra and P.J.E. Peebles, Phys. Rev. {\bf D7}, 3406 (1988). 
\bibitem{Wetterich 88} C. Wetterich, Nucl. Phys. {\bf B302}, 668  (1988). 
\bibitem{WCL 93} D. Wands, E,J, Copeland and A.R. Liddle, Ann. N. Y. Acad. Sci. {\bf 688}, 647 (1993). 
\bibitem{Ferreira+Joyce 97} P.G. Ferreira and M. Joyce, Phys. Rev. Lett. 79, 4740 (1997) . 
\bibitem{CLW 98} E.J. Copeland, A.R. Liddle and D. Wands, Phys. Rev. {\bf D57}, 4686  (1998). 
\bibitem{Coble et al 97} K. Coble, S. Dodelson and J. A. Frieman, Phys. Rev. {\bf D55}, 1851  (1997). 
\bibitem{Wetterich 95} C. Wetterich, Astron. Astrophys. {\bf 301}, 321 (1995) . 
\bibitem{Viana+Liddle 98} P.T.P. Viana and A.R. Liddle, Phys. Rev. {\bf D57}, 674 (1998). 
\bibitem{Liddle+Scherrer 99} A.R. Liddle and R.J. Scherrer, Phys. Rev. {\bf D59}, 023509 (1999). 
\bibitem{Billyard+Coley+Hoogen 98} A.P. Billyard, A.A. Coley and R.J. van den Hoogen, Phys. Rev. {\bf D58}, 123501 (1998). 
\bibitem{Hoogen+Coley+Wands 99}  R.J. van den Hoogen, A.A. Coley and D. Wands, Class. Quantum Grav.  {\bf 16},  1843 (1999). 
\bibitem{Coley+Ibanez+Hoogen 97} A.A. Coley, J. Iba\~nez and R.J. van den Hoogen, J. Math. Phys. {\bf 38},  525 (1997). 
\bibitem{Billyard+Coley+Ibanez 98} A. Billyard, A.A. Coley and J. Iba\~nez, Phys. Rev. {\bf D59}, 023507 (1999). 
\bibitem{Uzan 99} J.P. Uzan, Phys. Rev. {\bf D59},  123510 (1999). 
\bibitem{Amendola 99} L. Amendola, Phys. Rev. {\bf D60}, 043501 (1999). 
\bibitem{Holden+Wands 00} D. Holden and D. Wands, Phys. Rev. {\bf D61},  043506 (2000). 
\bibitem{Nunes+Mimoso 00b} A. Nunes and J.P. Mimoso, Phys. Lett. {\bf B488}, 423 (2000). 
\bibitem{Billyard+Coley 00} A. Billyard and A.A. Coley, Phys. Rev. {\bf D61}, 083503  (2000). 
\bibitem{Barrow 86} J.D. Barrow, Phys. Lett. {\bf B180}, 335 (1986). 
 \bibitem{Lima+Germano 92} J.A.S. Lima and A.S.M. Germano, Phys. Letts. {\bf A170}, 373  (1992). 
\bibitem{Zimdahl+Triginer+Pavon 96} W. Zimdahl, J. Triginer and D. Pav\'on, Phys. Rev. {\bf D54}, 6101 (1996). 
\bibitem{Chimento+Jakubi+Pavon 99} L.~P.~Chimento, A.~S.~Jakubi and D.~Pavon, Int.\ J.\ Mod.\ Phys.\  {\bf D9}, 43 (2000). 
\bibitem{Barrow 88} J.D. Barrow, Nuc. Phys. {\bf B310}, 743  (1988). 
\bibitem{Madsen et al 92} M.S. Madsen, J.P. Mimoso, J. Butcher and G.F.R. Ellis, Phys. Rev. {\bf D46}, 1399 (1992). 
\bibitem{Wald 83} R. M. Wald, Phys. Rev. {\bf D28}, 2118 (1983).
\bibitem{Coley+Wainwright 92} A.A. Coley and J. Wainwright, Clas. Quantum Grav. {\bf 9}, 651 (1992). 
\bibitem{Lucchin+Matarrese 85} F. Lucchin and S. Matarrese, Phys. Rev. {\bf D 32}, 1316 (1985).  
\bibitem{Halliwell 87} J.J. Halliwell, Phys. Lett. {\bf B185}, 341 (1987).  
\bibitem{Burd+Barrow 88} A.B. Burd and J.D. Barrow, Nucl. Phys. {\bf B308}, 929 (1988). 
  
\bibitem{Bergmann:1968} P. G. Bergmann, Int. J. Theor. Phys., {\bf 1} (1968), 25.  
\bibitem{Wagoner:1970} R. V. Wagoner, Phys. Rev. {\bf D1} (1970), 3209.  
\bibitem{Nordtvedt:1970} K. Nordtvedt, Astrophys. J., {\bf 161} (1970), 1059.  
\bibitem{Will 93} C.M. Will, {\em Theory and Experiment in Gravitation}, Cambridge University Press (1993).  
\bibitem{B+D 61} C. Brans and R.H. Dicke, Phys. Rev. {\bf D124},  925 (1961). 
\bibitem{Dicke 62} R.H. Dicke, {\sl Phys. Rev.} {\bf 125}, 2163 (1962). 
\bibitem{Mimoso+Wands 95} J.P. Mimoso and D. Wands, Phys. Rev. {\bf D51}, 477 (1995).  
\bibitem{Mimoso+Nunes 98} J.P. Mimoso and A. Nunes, Phys. Lett. {\bf A248}, 325 (1998). 
\bibitem{Amendola 00} L.~Amendola, Phys.\ Rev.\  {\bf D62}, 043511 (2000). 
\bibitem{Nariai 68} H. Nariai, Prog. Theor. Phys. {\bf 40}, 49 (1968). 
\bibitem{Mimoso 93} J.P. Mimoso, D-Phil. Thesis, University of Sussex (1993). 
\bibitem{Barrow+Mimoso 94} J.D. Barrow and J.P. Mimoso, Phys. Rev. {\bf D50}, 3746 (1994). 
\bibitem{Bernstein 88} J. Bernstein, {\em Kinetic Theory of the expanding universe}, Cambridge University Press (1988).
\bibitem{White et al 93} S.D.M. White, J.F. Navarro, A.E. Evrad and C. S. Frenk, Nature {\bf 366}, 429 (1993). 
\bibitem{Perlmut et al 98} S. Perlmutter et al, Nature {\bf 391}, 51 (1998). 
\bibitem{Efstathiou et al 99} G. Efstathiou et al., Mon. Not. R. Astron. Soc. {\bf 303}, L47 (1999). 
\bibitem{Weinberg 72} S. Weinberg, {\em Gravitation and Cosmology}, John Wiley and Sons, New York (1972). 
\bibitem{Kolb+Turner 90} E.W. Kolb and M.S. Turner, {\em The Early Universe}, Addison-Wesley Publishng Company, Redwood City, California (1990).  
\bibitem{B+K 77} V. A. Belinskii and I. M. Khalatnikov, Sov. Phys. JETP {\bf 45}, 1 (1977)
 
\end{thebibliography}
\end{document}